\documentclass[a4paper,11pt]{article}
\usepackage[utf8]{inputenc}

\usepackage{geometry}
\usepackage{setspace}
\usepackage{authblk}

\usepackage{array, threeparttable, booktabs,caption}
\usepackage{times}
\usepackage{enumitem}
\usepackage{hyperref}
\usepackage{tabularx}
\usepackage{graphicx}
\newcommand{\RNum}[1]{\uppercase\expandafter{\romannumeral #1\relax}}
\newcommand{\beginsupplement}{%
        \setcounter{section}{0}
        \renewcommand{\thesection}{S\arabic{section}}
        \setcounter{table}{0}
        \renewcommand{\thetable}{S\arabic{table}}%
        \setcounter{figure}{0}
        \renewcommand{\thefigure}{S\arabic{figure}}%
     }

\begin{document}

\title{\textbf{Interdisciplinary researchers attain better performance in funding}}
\author[1]{Ye Sun}
\author[2,*]{Giacomo Livan}
\author[3]{Athen Ma}
\author[1,4,5,6]{Vito Latora}
\affil[1]{School of Mathematical Sciences, Queen Mary University of London, London E1 4NS, UK}
\affil[2]{Department of Computer Science, University College London, London WC1E 6EA, UK.}
\affil[3]{School of Electronic Engineering and Computer Science, Queen Mary University of London, London E1 4NS, UK}
\affil[4]{Dipartimento di Fisica ed Astronomia, Universit\`a di Catania and INFN, 95123, Catania, Italy}
\affil[5]{The Alan Turing Institute, The British Library, London NW1 2DB, UK}
\affil[6]{Complexity Science Hub Vienna (CSHV), Vienna, Austria}

\affil[*]{To whom correspondence and requests for materials should be addressed. E-mail: g.livan@ucl.ac.uk}
\date{}

\maketitle

\begin{abstract}
    Interdisciplinary research is fundamental when it comes to tackling complex problems in our highly interlinked world, and is on the rise globally. Yet, it is unclear why — in an increasingly competitive academic environment — one should pursue an interdisciplinary career given its recent negative press. Several studies have indeed shown that interdisciplinary research often achieves lower impact compared to more specialized work, and is less likely to attract funding. We seek to reconcile such evidence by analyzing a dataset of $44,419$ research grants awarded between $2006$ and $2018$ from the seven national research councils in the UK. We compared the research performance of researchers with an interdisciplinary funding track record with those who have a specialized profile. We found that the former dominates the network of academic collaborations, both in terms of centrality and knowledge brokerage; but such a competitive advantage does not immediately translate into impact. Indeed, by means of a matched pair experimental design, we found that researchers who transcend between disciplines on average achieve lower impacts in their publications than the subject specialists in the short run, but eventually outperform them in funding performance, both in terms of volume and value. Our results suggest that launching an interdisciplinary career may require more time and persistence to overcome extra challenges, but can pave the way for a more successful endeavour. 
\end{abstract}

\section*{Introduction}
Interdisciplinary research is increasingly regarded as the key to tackle contemporary complex societal challenges and to stimulate scientific innovation~\cite{ledford2015solve,rylance2015grant,sun2020evolution}. As high-impact discoveries often occur at the intersection of disciplines~\cite{szell2018nobel,thurner2020role}, scientists have become more engaged in research areas that transcend the boundaries between traditional fields~\cite{van2015interdisciplinary,ma2018scientific}, and increasingly collaborate across such boundaries~\cite{brown2015interdisciplinarity}.

A higher uptake in interdisciplinary research has been widely reported across academia~\cite{battiston2019taking}, as there is now a much greater level of knowledge transfer between subjects among researchers~\cite{zeng2019increasing}. Yet, these trends are somewhat intriguing when one looks at the evidences available on research \emph{outcomes}, which suggest that -- more often than not -- interdisciplinary research may be an unrewarding enterprise in today's highly competitive academic environment. In fact, it is only interdisciplinary work based on proximal combinations of different fields that achieves recognition and impact (as quantified by accrued citations), whereas distal combinations are usually perceived as too risky or heterodox~\cite{yegros2015does}. Similarly, interdisciplinary research is often associated with lower citation rates~\cite{levitt2008multidisciplinary,elsevier2015review} together with lower funding success~\cite{bromham2016interdisciplinary}.

With such seemingly gloomy prospects on interdisciplinary research, we asked if those who pursue this line of work share the same fate? Namely, we compared the career progressions of researchers with a track record of research funding that can be unambiguously classified as either interdisciplinary or monodisciplinary. We examined data detailing more than $44,000$ research grants funded between $2006$ and $2018$ by the seven discipline-based UK national research councils (collectively form the bulk of the largest public funding body in the UK), which provide funding to universities and academic institutions to undertake research across a broad spectrum of fields, including arts and humanities, biology, economics, engineering and physics, medicine, environmental sciences and astronomy (Table~\ref{tab:CouncilProperty}). 

Through network analysis, we discovered that researchers who study across different disciplines play a crucial role of knowledge brokers in the academic collaboration network, bridging the gap between subjects and researchers that may otherwise remain disconnected. Such a role pays off in the long run. By means of a matched pair experimental design we found that, despite achieving comparatively lower impact, interdisciplinary researchers outperform their discipline-specific peers in funding performance, both in terms of the number of grants and their funding size. Our findings help explain the continuous drive on interdisciplinary research, and provide insights on its role in the modern research funding landscape that may be useful to researchers and funding bodies alike.

\section*{Evolution of cross-council behaviors}
\begin{figure}
\centering
    \includegraphics[width=16cm]{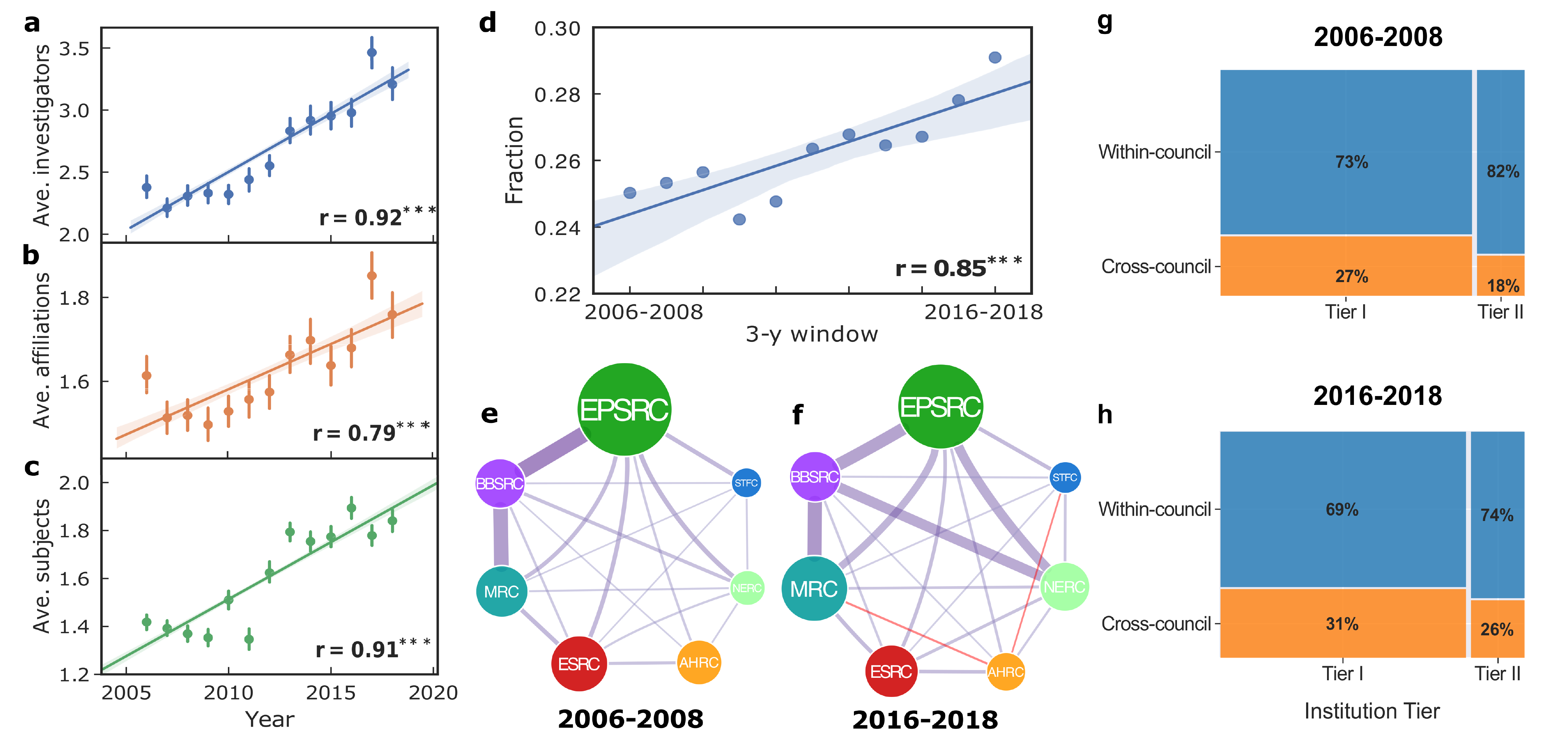}
    \caption{\textbf{Time evolution of the funding landscape.} \textbf{(a)} The typical number of team members per grant shows a significant increase over time. \textbf{(b)} The average number of affiliations participating in each grant grows with time. \textbf{(c)} The average number of subjects listed in each grant continues to rise over time. \textbf{(d)} The fraction of cross-council investigators increases over time (regardless of the length of the sliding window or investigator groups; see Fig.~\ref{fig:TestWindow} and Fig.~\ref{fig:TestNumber}).
    In panels \textbf{(a)} to \textbf{(d)}, the solid line and the shaded area represent the regression line and the $95\%$ confidence intervals, respectively. \textbf{(e-f)} The co-activity network of investigators in  two time windows, 2006-2008 and 2016-2018. The color from each of the research council's logo has been used to denote each node. Two new links appeared in 2016-2018 are highlighted in red. \textbf{(g-h)} The percentage of cross-council investigators in different institutional tiers and periods. Here, the research institutions are stratified into two tiers by checking whether their total awarded funding is larger than the average amount per institution (i.e., $\textsterling 1.02 \times 10^8$). Box widths are proportional to the number of investigators in Tier I and Tier II, respectively. Box heights are proportional to the percentage of cross-council and within-council investigators. The institutions in Tier I have a higher proportion of cross-council investigators than those in Tier II in both 3-y time windows ($\chi^2$ test $p < 0.0001$, odds ratio$ = 1.67$ for 2006-2008; $p < 0.0001$, odds ratio$ = 1.28$ for 2016-2018). The same conclusions have been obtained when different time window lengths and different criteria of institutional stratification have been used (see Fig.~\ref{fig:DifferenceTier2_5y} and Fig.~\ref{fig:DifferenceRussell_3y}).
    \label{fig:EvolutionGrant}}
\end{figure}

Between $2006$ and $2018$, the average team size increased over time, with team composition becoming more cross-institutional (Fig.~\ref{fig:EvolutionGrant}a-b). This demonstrates an increasing trend of collaborative science in the UK funding landscape, which is consistent with rising teamwork and multi-institutional research in scientific publications~\cite{guimera2005team,wuchty2007increasing,jones2008multi}. A funded project can be associated with one or more research subjects out of $104$ possible subjects. The level of cross-disciplinarity showed an upward trajectory (Fig.~\ref{fig:EvolutionGrant}c), with nearly half of funded projects ($44\%$) being related to at least two research subjects. This finding is consistent with the general shift towards more cross-disciplinary research~\cite{ledford2015solve}.

The increased inter-linkage among different research subjects in funded projects implies that some research projects may fall into the remits of more than a single research council, which can potentially lead to more funding opportunities for researchers who perform cross-disciplinary research. To examine this, we divided the investigators into two groups: cross-council investigators and within-council investigators. The former are those who have obtained funding from at least two different research councils; while the latter are those who have received funding from one research council only. A sliding window of three years was used to examine how the two groups of investigators evolved over time. Starting with $2006$, we shifted the sliding window year by year and obtained a total of $11$ time slices. As expected, the fraction of cross-council investigators shows a marked increase, from around $0.24$ in 2006-2008 to $0.29$ in 2016-2018 (Fig.~\ref{fig:EvolutionGrant}d).

To better understand how this rise in cross-council investigators alters the funding landscape, we constructed a co-activity network whereby nodes represent research councils. Two councils share a link if they have supported the same investigator, and a link is weighted by the number of investigators who have obtained grants from both the connected councils. Starting with the 2006-2008 window (Fig.~\ref{fig:EvolutionGrant}e), cross-council investigators are most commonly found between \textsl{BBSRC} and \textsl{EPSRC}, and between \textsl{BBSRC} and \textsl{MRC}, while the numbers are noticeably lower between other research councils. A decade later (2016-2018 window), the co-activity network becomes fully connected (Fig.~\ref{fig:EvolutionGrant}f) with two new links connecting \textsl{AHRC} with \textsl{MRC}, and \textsl{STFC}, respectively. In addition, the numbers of cross-council investigators between \textsl{BBSRC} and \textsl{NERC}, and between \textsl{EPSRC} and \textsl{NERC} soar by $39\%$ and $142\%$, respectively. These shifts in the funding landscape appear to be the response to the \textsl{UKRI} policy to support research across council boundaries and enhance the culture of multidisciplinary research~\cite{stokstad2018research}.

Elite institutions have been found to be prime recipients of research funding, as they are key in orchestrating collaborations ~\cite{torres2020not} and generating research outputs ~\cite{li2019early,way2019productivity}. We considered the total amount of funding received by institutions between $2006$ and $2018$ as a proxy of their national rank, and examined the level of cross-council activities among their investigators. For the sake of simplicity, institutions are grouped into two tiers, with the top tier consisting of $40$ institutions (Tier \RNum{1}) that have received a higher than average total funding over the aforementioned $13$ years period, and the remaining institutions forming the bottom tier (Tier \RNum{2}). There are noticeably more cross-council investigators in top tier institutions (Fig.~\ref{fig:EvolutionGrant}g-h and Fig.~\ref{fig:InstitutionFeature}), in line with previous findings on the governing role on research innovation among top institutions~\cite{ma2015anatomy}. Interestingly, the proportion of cross-council investigators in the bottom tier shows a bigger increase, from $18\%$ to $26\%$, which is twice of the top tier (with an increase from $27\%$ to $31\%$). 

\section*{Structural advantage in the collaboration network}
Our results have shown that interdisciplinary research is undoubtedly gaining momentum, with more cross-council investigators emerging across the University sector. To better understand this shift in collaborative science, we examined research partnerships among investigators and studied the roles played by the cross-council and within-council groups. Here, network nodes are the investigators, and two investigators are connected if they have partnered in one or more research projects. 

\begin{figure}
\centering
    \includegraphics[width=17cm]{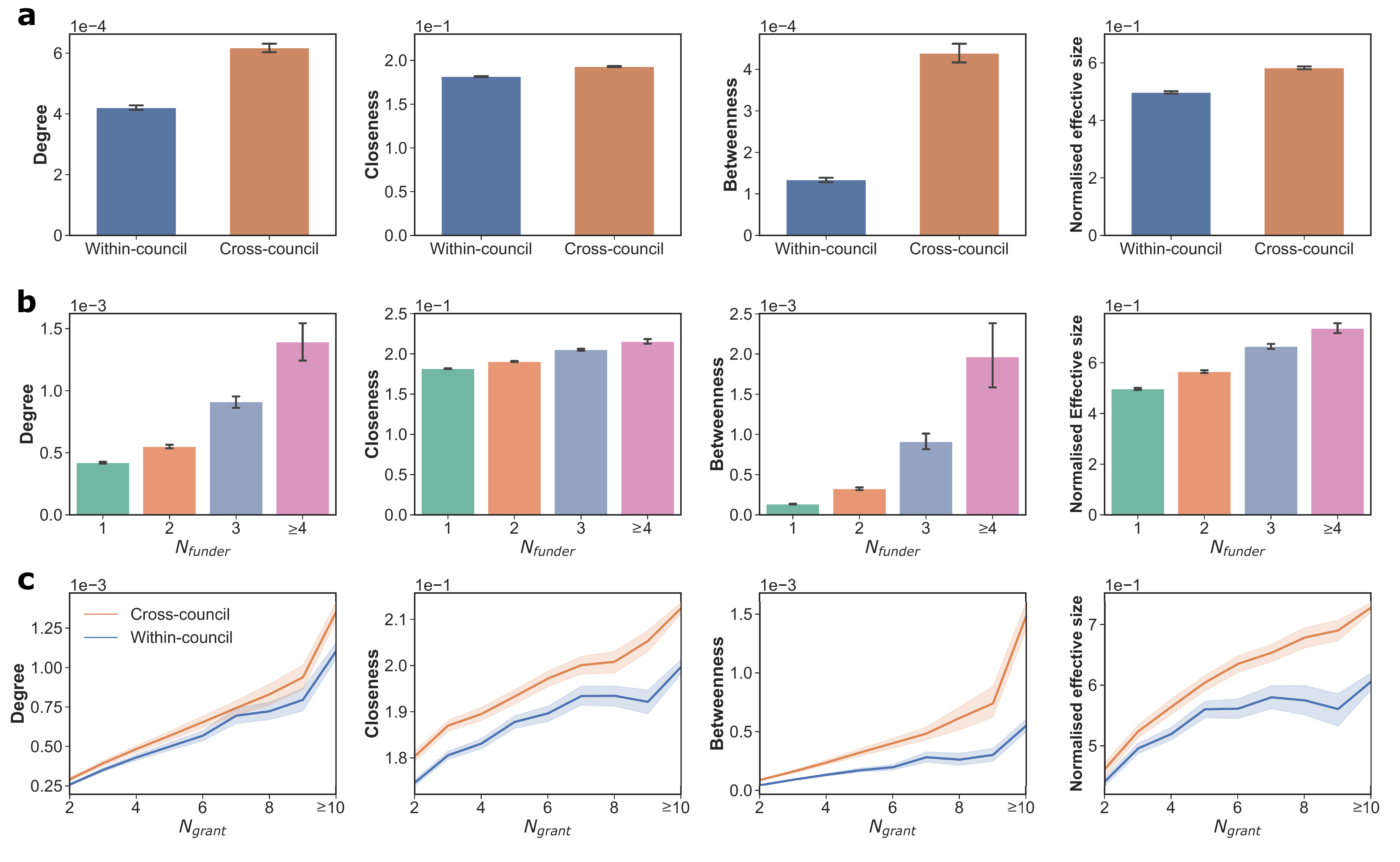}
    \caption{\textbf{Structural advantage of cross-council investigators in the collaboration network.} Each column corresponds to a different network property, namely: degree centrality, closeness centrality, betweenness centrality and normalized effective size. \textbf{(a)} Cross-council investigators significantly outperform the within-council investigators in all four network properties (Welch’s t-test, $p<0.001$ in all cases). \textbf{(b)} Network metrics among investigators increase with the number of councils they have received funding from. \textbf{(c)} Cross-council investigators consistently have a network advantage over within-council investigators with reference to degree, closeness, betweenness and normalized effective size. The error bars and shaded areas represent $95\%$ confidence intervals.}
    \label{fig:NetworkCompare}
\end{figure}

Cross-council investigators consistently show a much broader collaborative practice with a much higher average degree (Fig.\ref{fig:NetworkCompare}a). They are also more likely to occupy prime locations or gateways for information dissemination, as demonstrated by both a higher closeness and betweenness centrality. Indeed, we found that cross-council investigators are much more likely to be brokers of information -- as they are characterized by a higher average effective network size~\cite{latora2013social} -- suggesting they play a central role in establishing partnerships. Overall, the more diversified their funding source, the more advantageous their network position appears to be (Fig.\ref{fig:NetworkCompare}b). By comparing the two groups of investigators with respect to their number of grants, our results on network metrics show systematic differences in their collaboration patterns (Fig.\ref{fig:NetworkCompare}c), and the differences are most apparent among the more successful investigators. 

\section*{Research outcomes and scientific impact}
\begin{figure}
\centering
    \includegraphics[width=14cm]{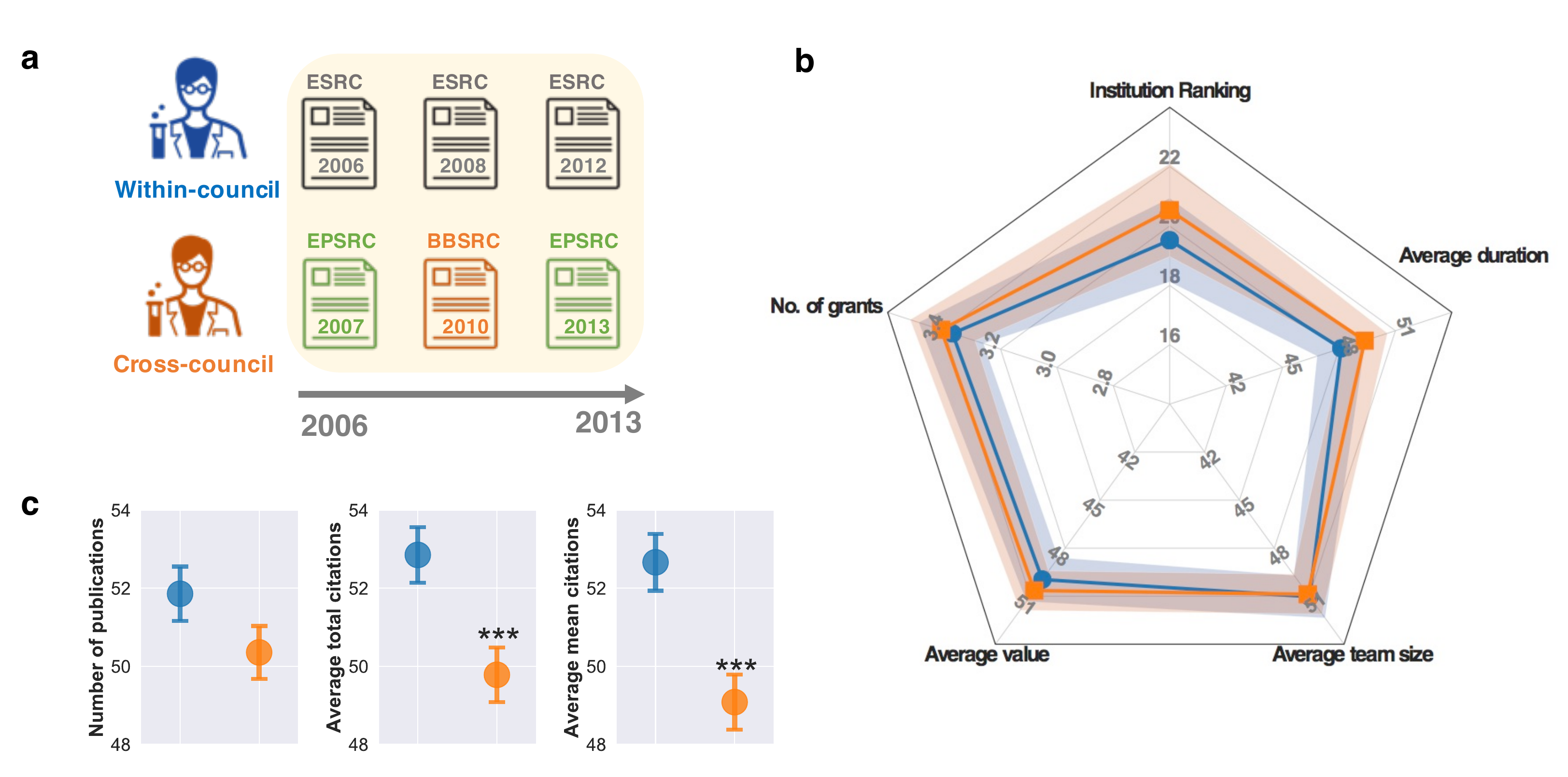}
    \caption{\textbf{Comparing scientific outcomes between cross-council and within-council investigators.} \textbf{(a)} An illustrative example of cross-council (orange) and within-council (blue) PIs within the observation window from $2006$ to $2013$. \textbf{(b)} Matching the cross-council and within-council PIs with similar career profiles in terms of funding performance. We matched $5$ different characteristics for PIs: institutional ranking of a given PI whereby institutions are ranked by their total amount of funding between $2006$ and $2018$, the number of grants a given PI has received, and among these projects, the average grant value, the average team size, and the average project duration. There is no statistically significant difference between the two groups of PIs across the five dimensions following the pairing. The shaded areas represent 95\% confidence intervals. \textbf{(c)} Differences in research outcomes between cross-council and within council PIs on the average number of papers reported per project, the average number of \emph{total} citations received per grant (calculated as the average of the total citations received by papers associated with a grant), and the average number of citations received per paper per grant (calculated first as the average of the citations received by papers associated with a grant , and then averaged over the total number of grants awarded to a PI). Citations are considered within $5$ years after publication. All dimensions considered in panels \textbf{(b)} and \textbf{(c)} (with the exception of institutional ranking and number of grants) are quantified by calculating their percentile rank in the same council and year. The significance levels shown refer to t-tests and Kruskal–Wallis tests. ***$p < 0.01$, **$p < 0.05$, *$p < 0.1$. Error bars represent the standard error of the mean.}
    \label{fig:ResearchOutcome}
\end{figure}

\begin{figure*}
\centering
    \includegraphics[width=16cm]{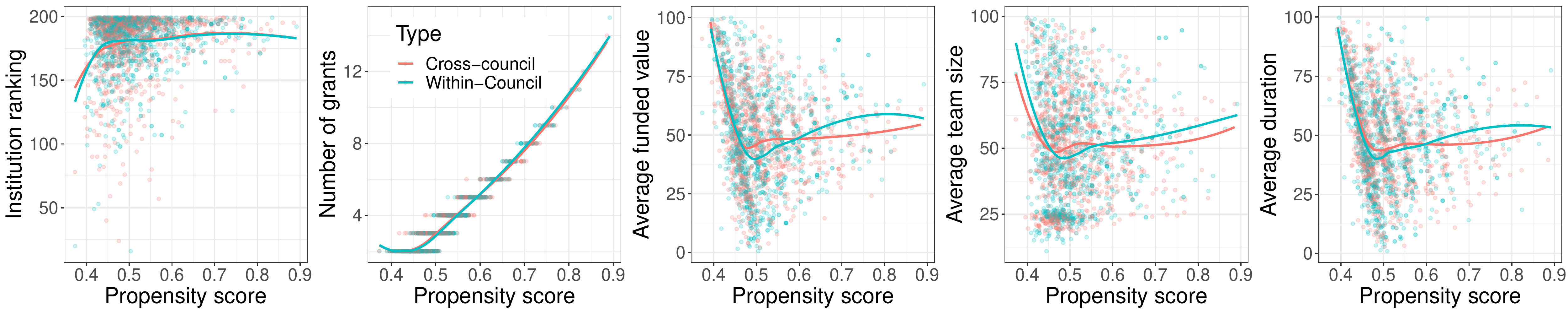}
    \caption{\textbf{Propensity scores for matched-pair analysis.} Each plot shows the propensity scores for each PI in our matched pair analysis for the five considered confounding factors (from left to right): (1) institutional ranking; (2) the number of awarded grants; (3) the average funded value per grant; (4) the average team size across the awarded grants; and (5) the average duration per project. Values for cross-council investigators are shown in red, while that of the within-council investigators are shown in green. Solid lines show the average values of all covariates as functions of the propensity scores.}
    \label{fig:DistributionPSM1}
\end{figure*}

Is there a detectable difference between the cross-council and within-council investigators in terms of research outcomes and scientific impact? To address this question, we needed to control for the bias caused by confounding factors identified, so that the observed differences in research outcomes and scientific impact between the two groups can be more confidently ascribed to interdisciplinarity (i.e., cross-council funding behaviour). 
Here, we performed a propensity score matching analysis ~\cite{ho2007matching} whereby an investigator's career profile is characterised by five confounding factors, namely the institutional ranking (measured by the total amount of funding received by the PI's institution), the number of grants awarded to a given PI, and their average funding value per grant, team size and project duration. The last three factors have been adjusted to account for variations in values in different disciplines, and over time (Section~\ref{sec:grant_pi_feature} and Fig.~\ref{fig:DifferenceCouncil}). A cross-council investigator is then paired with a within-council investigator if the two share a comparable career profile between $2006$ and $2013$ (Fig.~\ref{fig:ResearchOutcome}a), thereby eliminating the effect of these confounding factors on the phenomena under investigation (Fig.~\ref{fig:ResearchOutcome}b and Fig.~\ref{fig:DistributionPSM1}). The analysis yields a total of $958$ pairs of cross-council and within-council PIs. 

For each matched pair of PIs, we compared their research performance based on the achievements reported in their grants awarded during $2006$ and $2013$ (but omitted projects that go beyond $2018$ as the achievements reported would be incomplete), including the average number of papers reported per project, the average number of \emph{total} citations received per grant, and the average number of citations received per paper per grant. Again, these metrics have been normalized across the different disciplines, and over time (Section~\ref{sec:grant_pi_feature}). We observed that while the two groups of investigators produced more or less the same number of publications, cross-council investigators clearly received less citations in general than their within-council counterparts (Fig.\ref{fig:ResearchOutcome}c middle and bottom and Table~\ref{tab:RegressionOutcome}), both in terms of total citations (\emph{t}-test, \emph{p}=$0.0021$) and mean citations (\emph{t}-test, \emph{p}=$0.0004$), which is consistent with findings in prior studies~\cite{van2015interdisciplinary,elsevier2015review}.

\section*{Long-term funding trajectory} 
\begin{figure}
\centering
    \includegraphics[width=15cm]{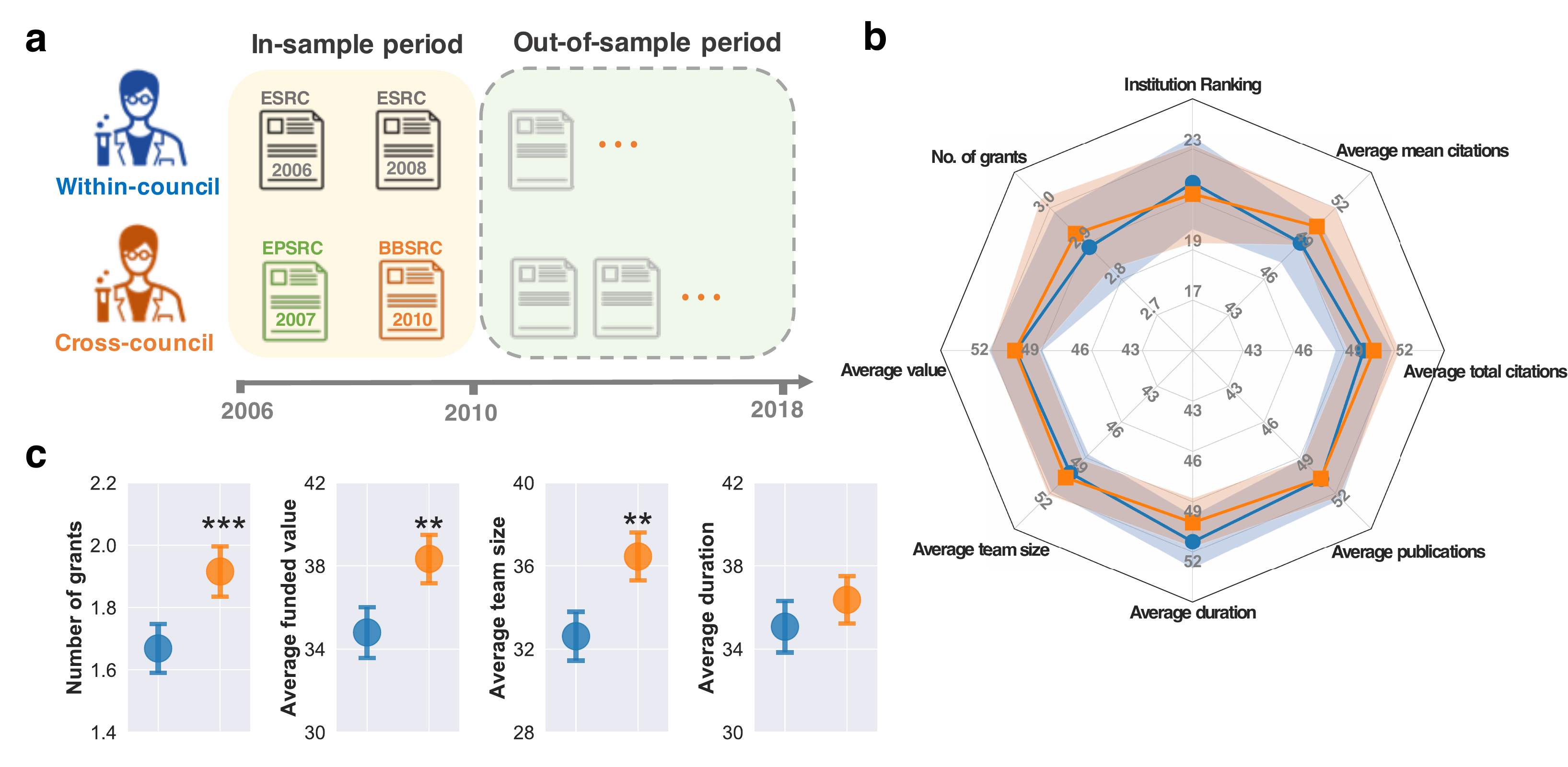}
    \caption{\textbf{Comparing long-term funding performance between cross-council and within-council investigators.} \textbf{(a)} An illustrative example of comparison in terms of long-term funding performance of cross-council (orange) and within-council (blue) PIs with similar funding profiles. \textbf{(b)} Matching the cross-council and within-council PIs with similar career profiles in terms of both funding performance and research outcomes during the in-sample period. We matched 8 different factors for PIs between $2006$ and $2010$ as follows: institutional ranking of a given PI whereby institutions are ranked by their total amount of funding between $2006$ and $2018$, the number of grants a given PI has received, and among these projects, the average grant value, the average team size, and the average project duration, the average number of publications reported , the average number of \emph{total} citations received per grant (calculated as the average of the total citations received by papers associated with a grant), and the average number of citations received per paper per grant (calculated first as the average of the citations received by papers associated with a grant , and then averaged over the total number of grants awarded to a PI). There is no statistically significant difference between the two groups of PIs across the eight factors following the pairing. The shaded areas represent the 95\% confidence interval. \textbf{(c)} Difference in long-term funding performance between cross-council and within-council PIs in the following eight years ($2011$ to $2018$). Cross-council PIs outperform within-council PIs in grant volume, value and team size. The significance levels shown refer to t-tests and Kruskal–Wallis tests. ***$p < 0.01$, **$p < 0.05$, *$p < 0.1$. Error bars represent the standard error of the mean.}
    \label{fig:FundingPerformance}
\end{figure}

\begin{figure*}
\centering
    \includegraphics[width=16cm]{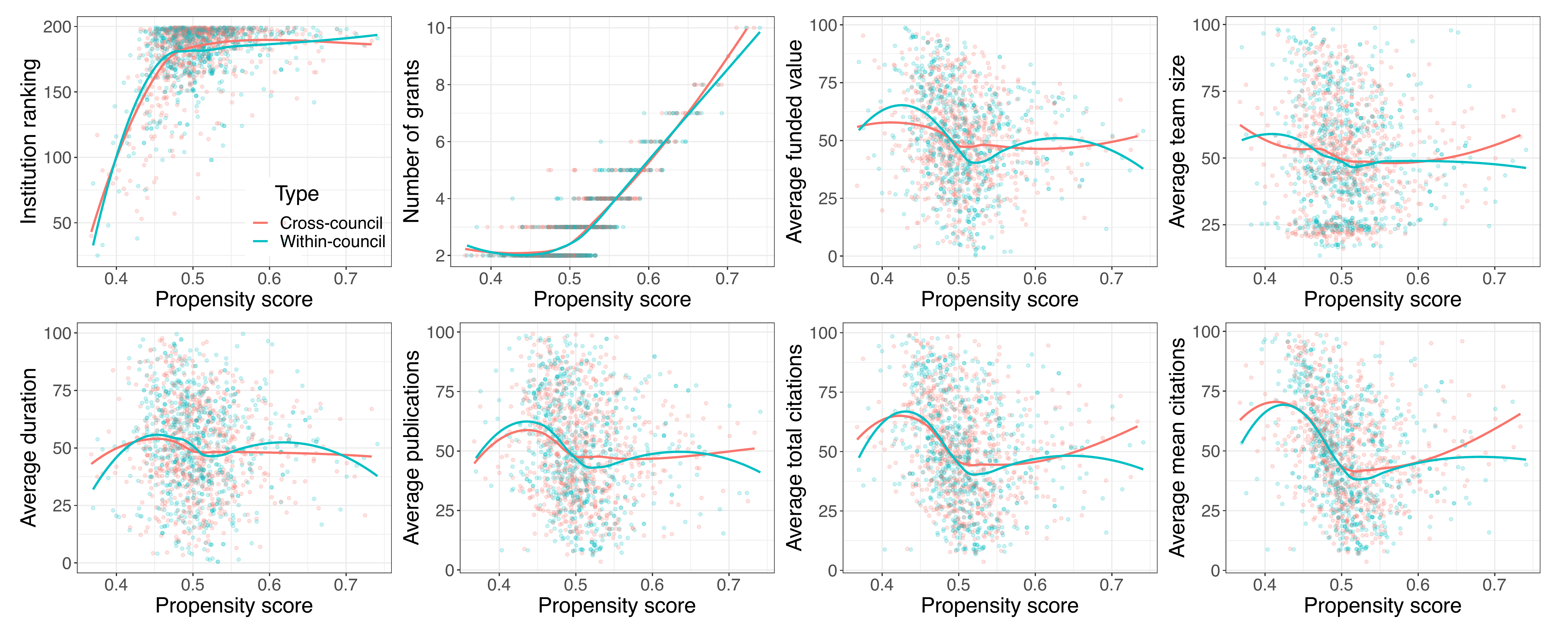}
    \caption{\textbf{Propensity scores for matched-pair analysis.} Each plot shows the propensity scores for each PI in our matched pair analysis for all considered confounding factors in the observation period, including (1) institutional ranking, measured by the total amount of research grants awarded to an institution between $2006$ and $2018$; (2) the number of grants awarded; (3) the average funded value per grant; (4) the average team size across the awarded grants; and (5) the average grant duration; (6) the average number of papers reported per project; (7) the average number of \emph{total} citations received per grant (calculated as the average of the total citations received by papers associated with a grant); and (8) the average number of citations received per paper per grant (calculated first as the average of the citations received by papers associated with a grant, and then averaged over the total number of grants awarded to a PI). Values for cross-council investigators are shown in red, while that of the within-council investigators are shown in green. Solid lines show the average values of all covariates as functions of the propensity scores.}
    \label{fig:DistributionPSM2}
\end{figure*}

\begin{figure*}
\centering
    \includegraphics[width=16cm]{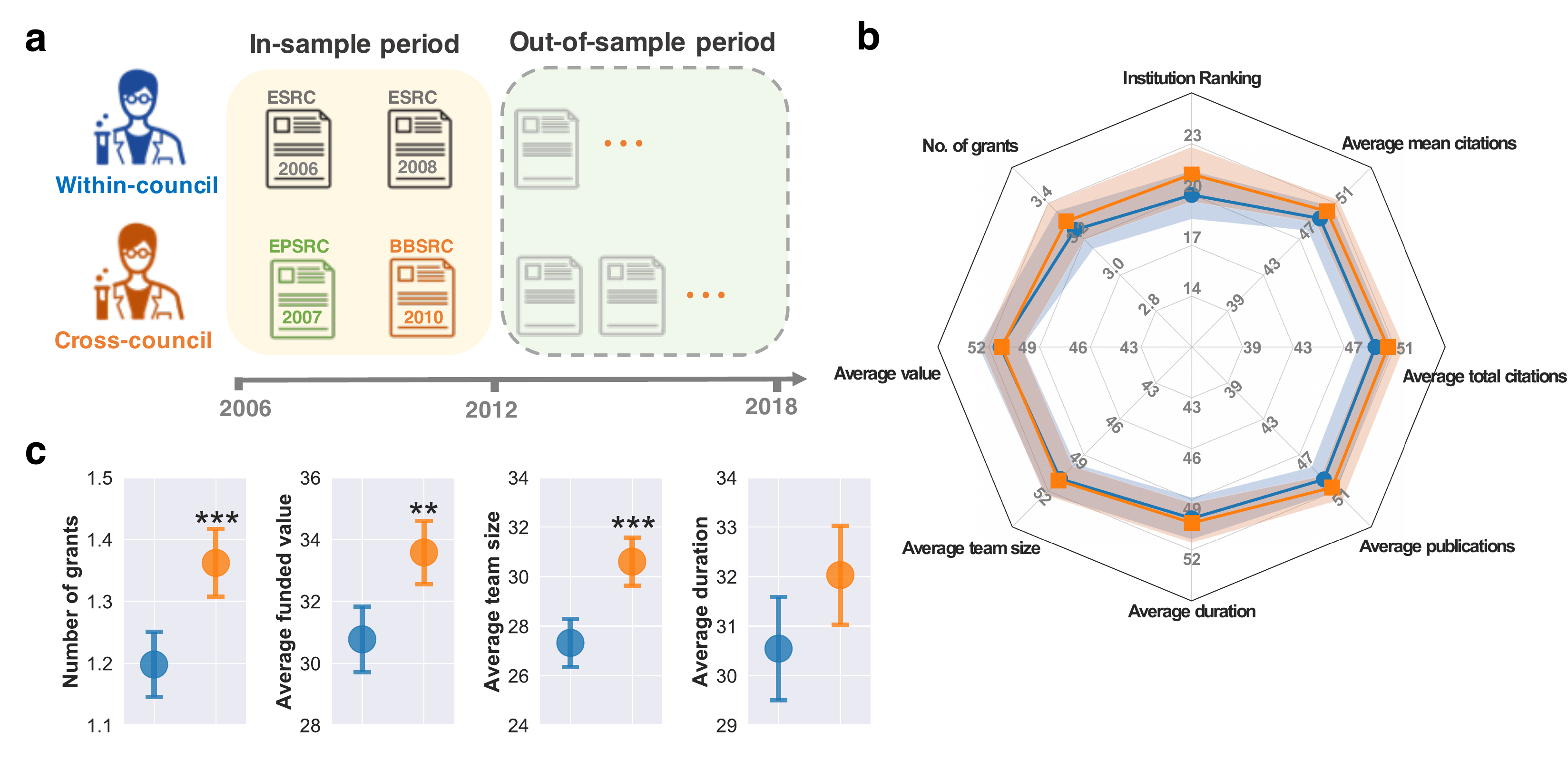}
    \caption{\textbf{Comparing long-term funding performance between cross-council and within-council investigators.} \textbf{(a)} An illustrative example of comparison in terms of future funding performance of cross-council (orange) and within-council (orange) PIs with similar funding profiles. Here, we consider the time window 2006-2012 as the in-sample period, and 2013-2018 as the out-of-sample period. \textbf{(b)} Matching the cross-council and within-council PIs with similar career profiles in terms of both funding performance and research outcomes during the observation period. We matched 8 different characteristics for PIs between $2006$ and $2012$ as follows: institutional ranking of a given PI whereby institutions are ranked by their total amount of funding between $2006$ and $2018$, the number of grants a given PI has received, and among these projects, the average grant value, the average team size, and the average project duration, the average number of publications reported , the average number of \emph{total} citations received per grant (calculated as the average of the total citations received by papers associated with a grant), and the average number of citations received per paper per grant (calculated first as the average of the citations received by papers associated with a grant , and then averaged over the total number of grants awarded to a PI). There is no statistically significant difference between the two groups of PIs across the eight factors following the pairing. \textbf{(c)} Difference in long-term funding performance between cross-council and within-council PIs in the subsequent $6$ years (from $2013$ to $2018$) . Cross-council PIs outperform within-council PIs in grant volume, value and team size. The significance levels shown refer to t-tests and Kruskal–Wallis tests. ***$p < 0.01$, **$p < 0.05$, *$p < 0.1$. Error bar represents the standard error of the mean.}
    \label{fig:FundingPerformance2}
\end{figure*}

\begin{figure*}
\centering
    \includegraphics[width=16cm]{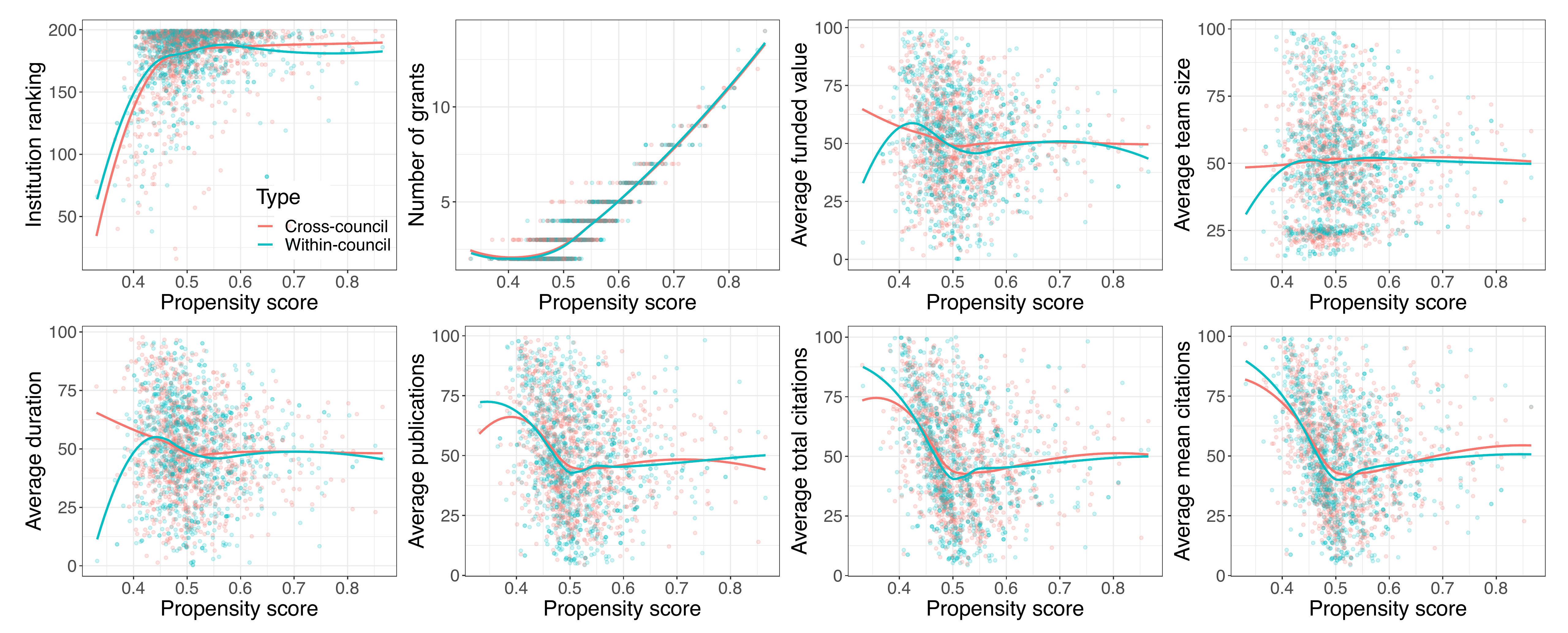}
    \caption{\textbf{Propensity scores for matched-pair analysis.} Each plot shows the propensity scores for each PI in our matched pair analysis for all considered confounding factors in the observation period, including (1) institutional ranking, measured by the total amount of research grants awarded to an institution between $2006$ and $2018$; (2) the number of grants awarded; (3) the average funded value per grant; (4) the average team size across the awarded grants; and (5) the average grant duration; (6) the average number of papers reported per project; (7) the average number of \emph{total} citations received per grant (calculated as the average of the total citations received by papers associated with a grant); and (8) the average number of citations received per paper per grant (calculated first as the average of the citations received by papers associated with a grant, and then averaged over the total number of grants awarded to a PI). Values for cross-council investigators are shown in red, while that of the within-council investigators are shown in green. Solid lines show the average values of all covariates as functions of the propensity scores.}
    \label{fig:DistributionPSM3}
\end{figure*}

We finally compared the funding trajectory of cross-council investigators to that of within-council investigators. We referred to 2006-2010 as the in-sample period where investigators are paired, and 2011-2018 as the out-of-sample period in which funding performance of each pair of investigators is compared (Fig.~\ref{fig:FundingPerformance}a). On this occasion, the pairing was done by not only matching their career profiles but also their research performance (i.e., reported achievements in grants described in the previous section, Fig.~\ref{fig:FundingPerformance}b and Fig.~\ref{fig:DistributionPSM2}), yielding $709$ investigator pairs. The cross-council investigator group outperformed their within-council counterparts, as demonstrated by the notable gains in the number of grants and their value; as well as the average team size (Fig.~\ref{fig:FundingPerformance}c, Table~\ref{tab:RegressionFuture}). For the sake of robustness, we repeated our analyses across different measurement time periods, and the same conclusions have been reached (Fig.~\ref{fig:FundingPerformance2} and Fig.~\ref{fig:DistributionPSM3}). These findings are important as they uncover previously unknown positive aspects of an interdisciplinary research career, providing a much needed optimistic outlook for those who wish to pursue this line of work~\cite{bromham2016interdisciplinary}.  

\section*{Discussion}
In this paper, we compared the careers of interdisciplinary investigators with those who are tied to a fixed discipline. In line with other findings on the rise of interdisciplinary research~\cite{battiston2019taking,zeng2019increasing}, we found that the fraction of cross-council investigators increased steadily during our period of observation. We also found that cross-council investigators sit more centrally than their peers in the academic collaboration network, which in turn provides them with considerable competitive advantage in terms of knowledge brokerage opportunities, but such a competitive advantage does not immediately merit a higher academic impact in their publications.

There are a number of possible reasons for the comparatively lower impact of projects led by cross-council investigators. It is reasonable to argue that their role as a knowledge broker leads to considerable costs -- both in terms of building collaborative relationship and establishing a common language to communicate across disciplines~\cite{goring2014improving} -- which may indirectly suppress their productivity. Also, despite a lack of consensus on its overall impact (see, e.g.,~\cite{lariviere2015long,leahey2017prominent}), a number of studies have shown that interdisciplinary research tends to garner recognition over longer periods of time compared to more specialized research~\cite{van2015interdisciplinary,wang2015interdisciplinarity}. In this respect, we ought to acknowledge that our results may partially be due to the duration covered by our available data, which constrained our analyses to quantifying impact with citations received within $5$ years of publication. It is plausible that our conclusions about impact may be different over longer time periods. This limitation notwithstanding, our findings suggest that it may be very challenging for a junior researcher to launch an independent interdisciplinary academic career. Indeed, all current practices of academic impact evaluation are to some extent influenced by citation-based bibliometric indicators~\cite{moher2018assessing} and, therefore, may be stacked against junior interdisciplinary researchers receiving citations at a slower pace (see, e.g.,~\cite{rafols2012journal}). 

However, and more importantly, the main result of our study shows that cross-council investigators eventually outperformed their peers in terms of funding. This result is robust and statistically significant across different dimensions, with respect to the number of funded grants, the average funded value and the average team size per grant awarded. Although at face value this may seem to contradict previous findings on the lower funding success rate of interdisciplinary research~\cite{bromham2016interdisciplinary}, we believe that this is not necessarily the case. Our results primarily focus on the funding performance of investigators in terms of \emph{volume and value} -- not of grant proposals -- and do not speak to their success \emph{rate}, as data about rejected proposals were not available to us. It is therefore possible that interdisciplinary investigators in our data may still encounter a lower success rate, although this would directly imply that they submit proposals in much larger numbers than their peers.

All in all, we believe that the more plausible explanation for our findings is that indeed interdisciplinary investigators develop -- on average -- a better ability to attract funding in the long run. Squaring this with their lower impact in the short run suggests that interdisciplinary investigators may be `late bloomers' who tend to achieve success over a longer period of time; however, there are indications that their more diversified research portfolios could give them an edge in securing long-term tenure~\cite{franzoni2017academic}.

\section*{Methods}
\subsection*{Data sets} 
We collected $44,419$ research projects conducted between $2006$ and $2018$ from \href{https://www.ukri.org/}{UK Research and Innovation (UKRI)}, which includes the grants information from seven national discipline-based research councils, namely AHRC, BBSRC, ESRC, EPSRC, MRC, NERC and STFC (see Section~\ref{sec:seven_council}). The basic information for each research council has been summarized in Table~\ref{tab:CouncilProperty}. The research grants cover the full spectrum of academic disciplines from the medical and biological sciences to astronomy, physics, chemistry and engineering, social sciences, economics, environmental sciences and the arts and humanities, which enables us to comprehensively investigate the research and innovation in the UK. For each research project, we recorded the information of the title, abstract, the start date and end date, principal investigator (PI) and co-investigators (CI), fund value, lead/collaborating institutions and scientific outcomes (i.e., publications). Grant was considered to be awarded to the PI and the affiliation of the PI. Information on how the overall funding of a given grant was divided among the rest of the investigators (and their affiliations) was not made available. Among them, there are $37,677$ research projects that have been classified with at least one research subject (a total of $104$ subjects). All the research projects, investigators and institutions have been assigned with unique IDs, which eliminates the problem of name disambiguation. 

For each research grant, all related papers published are recorded with the information of title and DOI. This provides the possibility for us to link the \textsl{UKRI} research grant database with the \href{https://www.microsoft.com/en-us/research/project/microsoft-academic-graph/}{Microsoft Academic Graph (MAG)} database by precisely matching the titles and DOI of the publications in two databases. MAG is a database consisting of large amount of scientific publications, their citation records, dates of publication, information regarding the authorship, publication venues and more. More importantly, the dataset specifies the keywords for each paper, as well as the position of each such keyword in a field-of-study hierarchy, the highest level of which is comprised of $19$ disciplines. Therefore, this connection between the two datasets not only offers us additional information about each paper, it also allows us to trace citations of each publication within the MAG and how these citations compare with other papers published in the same year and discipline. In the end, we matched a total of $409,546$ publications and calculated their accumulated citations 5 years after publication.

\subsection*{Evolution of cross-council behavior}
\begin{figure*}
\centering
    \includegraphics[width=10cm]{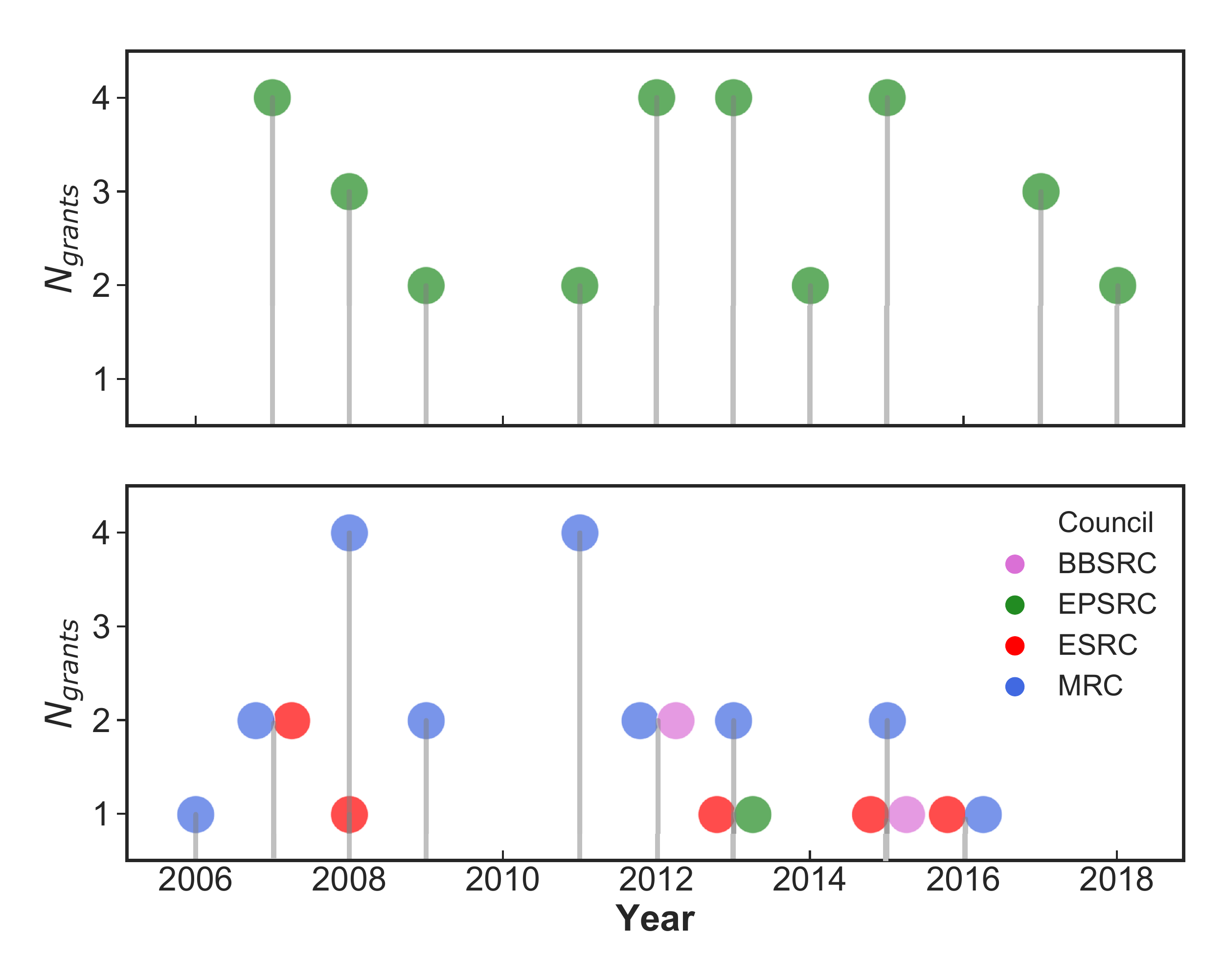}
    \caption{\textbf{The awarded grant history of two investigators.} Both investigators have obtained $30$ research grants throughout the studied period, but one has only received funding from \textit{EPSRC} (within-council; top panel) and the other has received funding from four different research councils (cross-council; bottom panel), namely \textsl{BBSRC}, \textsl{EPSRC}, \textsl{ESRC} and \textsl{MRC}. Here, the research projects in which the two investigators were involved as PIs and CIs have been documented.}
    \label{fig:concrete}
\end{figure*}

In the sliding window analysis, only investigators with at least two research grants have been tracked. We have excluded those with one research grant only as they will distort the number of within-council investigators. Fig.~\ref{fig:concrete} illustrates the grant history of a cross-council investigator and a within-council investigator. Although the two investigators have obtained the same number of research grants throughout the studied period, their funding trajectories from the research councils can be strikingly different.

\subsection*{Collaboration network}
The collaboration network was constructed by referring project partnerships between investigators between $2006$ and $2018$. In this network, nodes are investigators (PIs or CIs), and a link refers to a project partnership between two nodes. Research grants comprising only one investigator (i.e. only the PI) have been excluded from the network. We extracted the Largest Connected Component (LCC) of the collaboration network which consists of $86\%$ of the investigators.

We then performed a node-level network analysis on all the investigators who are in the LCC, and only included those with at least $2$ research grants during the studied period. In total, we obtained $6,911$ cross-council investigators and $12,563$ within-council investigators. To further test whether this structural advantage exists across different time periods, we examined the collaboration network in the first 5-y window (2006-2010) and last 5-y window (2014-2018) of the available period, and found that our conclusions remain unchanged (see Fig.\ref{fig:NetworkCompare2006} and Fig.\ref{fig:NetworkCompare2014}).

\subsection*{Normalized effective size}
The normalized effective size of node $i$’s ego network measures to which extent each of the first neighbours of $i$ is non-redundant with respect to the other neighbours. Formally, for the case of unweighted and undirected graphs, the normalized effective size of a node $i$ can be defined as~\cite{latora2013social}:
\begin{equation}
\centering
\zeta_{i} = 1-\frac{k_{i}-1}{k_i}C_{i} 
\end{equation}
where $k_{i}$ is the degree of the node and $C_{i}$ is its clustering coefficient. This indicator can vary from $0$ to $1$ with $\zeta_{i} = 0$ when the neighborhoods of $i$ are fully connected, and $\zeta_{i}$ taking its largest value $1$ when $i$ is the center of a star, and there are no links among its collaborators. Generally, the larger the value of $\zeta_{i}$, the less connected the neighborhood of $i$ is, and consequently, the higher the brokerage opportunities for investigator $i$. Investigators acting as brokers, on the one hand, tend to exhibit weak ties with their collaborators, on the other, are likely to gain exposure to a greater variance and novelty of information and link people with different ideas and perspectives~\cite{burt2009structural,ahuja2000collaboration}.

\subsection*{Propensity score matching}
To avoid the potential bias of covariates among PIs in the cross-council and within-council groups, we performed the propensity score matching analysis based on multi-variable logistic regression models, which is a statistical technique typically used to infer causality in observational studies~\cite{ho2007matching}. Propensity scores (PSs) are defined as the predicted probability of being a cross-council PI conditional upon a set of observed covariates. Cross-council PIs are matched to within-council PIs based on their PSs in one-to-one ratio, using a nearest-neighbor algorithm within a caliper of 0.01 on the probability scale. After the matching, the characteristics of cross-council and within-council groups in all observed covariates are statistically indistinguishable, with standardized differences $d < 0.1$, t-tests \textit{p}-value $>0.1$ for the sample means, and Kruskal–Wallis tests \textit{p}-value $>0.1$ for the entire distributions.

\bibliographystyle{naturemag}
\bibliography{scibib}

\section*{Data availability}
The UKRI funding data used in the paper are publicly accessible and can be downloaded via ~\url{https://www.ukri.org}. The publication and citation data are available via Microsoft Academic (\url{https://academic.microsoft.com}).

\section*{Code availability}
The code for used to perform pair matching is available at~\url{https://github.com/benmiroglio/pymatch}. All other codes used in this study are available from the corresponding author upon reasonable request.

\section*{Acknowledgements}
We thank Xiancheng Li for sharing the MAG data. V.L. acknowledges the support for this research from the Leverhulme Trust Research Fellowship “CREATE: the network components of creativity and success”.

\section*{Author contributions}
All authors designed research; Y.S. analyzed the empirical data and performed research; Y.S., G.L., A.M., and V.L. analysed data and wrote the paper.

\section*{Competing interests}
The authors declare that they have no competing interests.

\clearpage

\beginsupplement


\begin{titlepage}
    \begin{center}
        \huge
        Supplementary Information for:\\
        \huge
         Interdisciplinary researchers attain better performance in funding\\
        \vspace{0.5cm}
        \Large
        Ye Sun$^{1}$, Giacomo Livan$^{2,\ast}$, Athen Ma$^{3}$, Vito Latora$^{1,4,5,6}$\\
        
        \vspace{0.5cm}
        \large
        $^{1}$School of Mathematical Sciences, Queen Mary University of London, London E1 4NS, UK\\
        $^{2}$Department of Computer Science, University College London, London WC1E 6EA, UK\\
        $^{3}$School of Electronic Engineering and Computer Science, Queen Mary University of London, London E1 4NS, UK\\
        $^{4}$Dipartimento di Fisica ed Astronomia, Universit\`a di Catania and INFN, 95123, Catania, Italy\\
        $^{5}$The Alan Turing Institute, The British Library, London NW1 2DB, UK\\
        $^{6}$Complexity Science Hub Vienna (CSHV), Vienna, Austria\\

        \vspace{0.5cm}
        \large
        $^{\ast}$Corresponding author. Email: g.livan@ucl.ac.uk.
        \end{center}
\end{titlepage}

\tableofcontents

\clearpage
\section{Basic information for seven research councils}\label{sec:seven_council}
This study referred to the data of research grants from seven national discipline-based research councils, which cover almost the full spectrum of research subjects in UK:
\begin{enumerate}
\item Arts $\&$ Humanities Research Council (AHRC) funds independent research across a wide-range of Arts and Humanities subjects;
\item Biotechnology and Biological Sciences Research Council (BBSRC) funds research in plants, microbes, animals, tools and technology underpinning biological research;
\item Economic and Social Research Council (ESRC) funds research and training on social and economic issues;
\item Engineering and Physical Sciences Research Council (EPSRC) funds research and training in engineering and the physical sciences; 
\item Medical Research Council (MRC) supports research across the biomedical spectrum; 
\item Natural Environment Research Council (NERC) funds research, survey, training and knowledge transfer in the environmental sciences; \item Science and Technology Facilities Council (SFTC) operates large scale science and engineering research facilities, and funds research mainly in astronomy, particle physics, space science and nuclear physics. 
\end{enumerate}
Table.~\ref{tab:CouncilProperty} summarizes the basic information for each research council, including the number of awarded research grants, the total funded value, the number of investigators and institutions involved, and the number of research subjects. EPSRC is the largest funding body that provides the largest number and amount of research funding, while the smallest research funder is AHRC, providing less than one-tenth of EPSRC's funding. In addition, EPSRC, MRC and ESRC are the top three research councils that involve the largest number of investigators.

\section{Robustness check for increasing cross-council behavior} \label{sec:cross_trend}
\begin{figure*}
\centering
    \includegraphics[width=16cm]{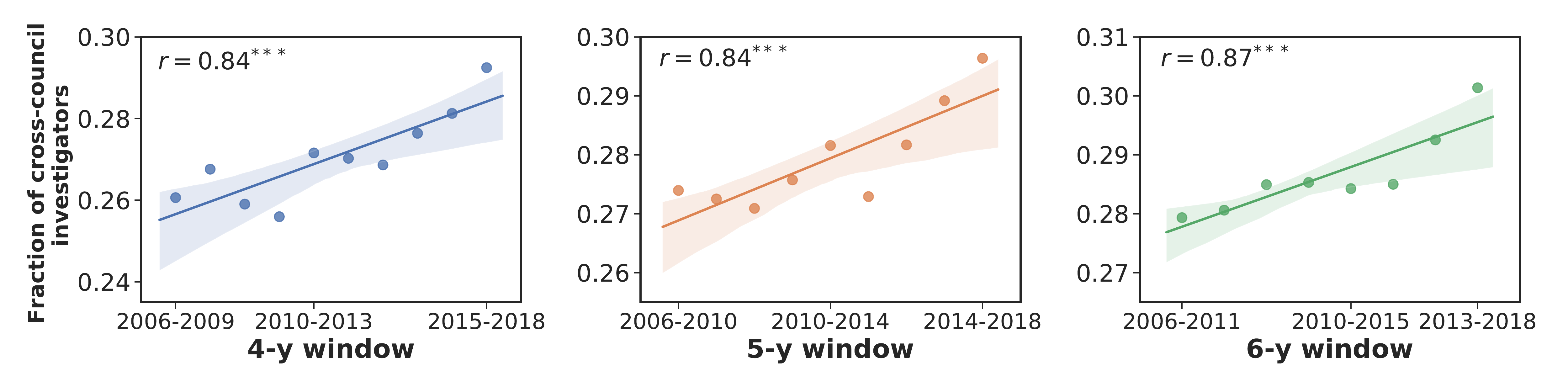}
    \caption{\textbf{Rising cross-council behaviour is observed regardless of the window size used.} The fraction of cross-council investigators shows a consistent increasing trend for different lengths of the sliding time window. Here, we consider investigators who have obtained at least $2$ research grants in each period. The solid line and the shaded area represent the regression line and the $95\%$ confidence interval, respectively. Each regression has also been annotated with the corresponding Pearson’s $r$ and $p$ values, where all $***p < 0.001$.}
    \label{fig:TestWindow}
\end{figure*}

To test whether our results only hold for 3-y sliding time window, we varied the lengths of time windows using 4-y, 5-y and 6-y. The increasing trend of cross-council behaviors is statistically significant regardless of the window lengths (Fig.~\ref{fig:TestWindow}).

\begin{figure*}
\centering
    \includegraphics[width=16cm]{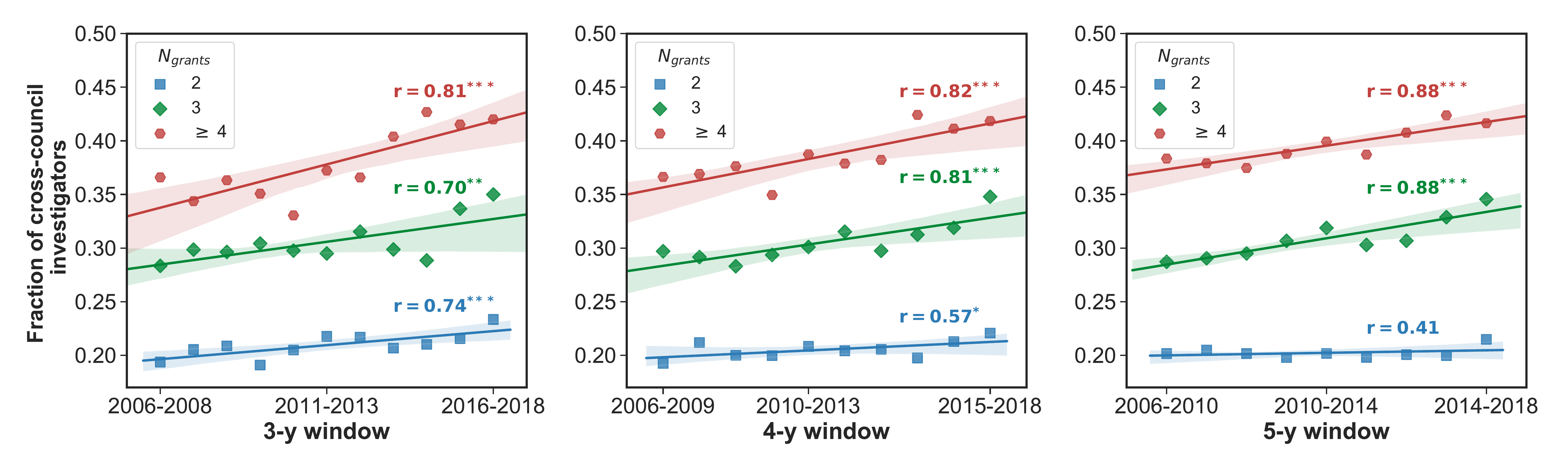}
    \caption{\textbf{Rising cross-council behaviour is observed among investigators with multiple grants.} Here, we first divide the investigators in each time window into three groups based on the number of research grants they have obtained, i.e., groups of investigators who have received two, three, or more than three research grants. We then calculate the fraction of cross-council investigators in each group and time window. The solid line and the shaded area represent the regression line and the $95\%$ confidence interval, respectively. Each regression has been annotated with the corresponding Pearson’s $r$ and $p$ values. ***$p < 0.01$, **$p < 0.05$, *$p < 0.1$.}
    \label{fig:TestNumber}
\end{figure*}

Furthermore, we also validated whether our results are robust when controlling for the investigators with similar number of research grants.  We separated the investigators in each time window into three groups, i.e., groups of investigators who received two, three, and more than three research grants, and then calculated the fraction of cross-council investigators in each group and time period. The fractions of cross-council investigators in each group show a similar rising pattern over time (Fig.~\ref{fig:TestNumber}), and the increase rates are slightly higher for the group of investigators who have obtained more number of research grants.

\section{Basic features of research institutions}
\label{sec:institution}
\begin{figure*}
\centering
    \includegraphics[width=16cm]{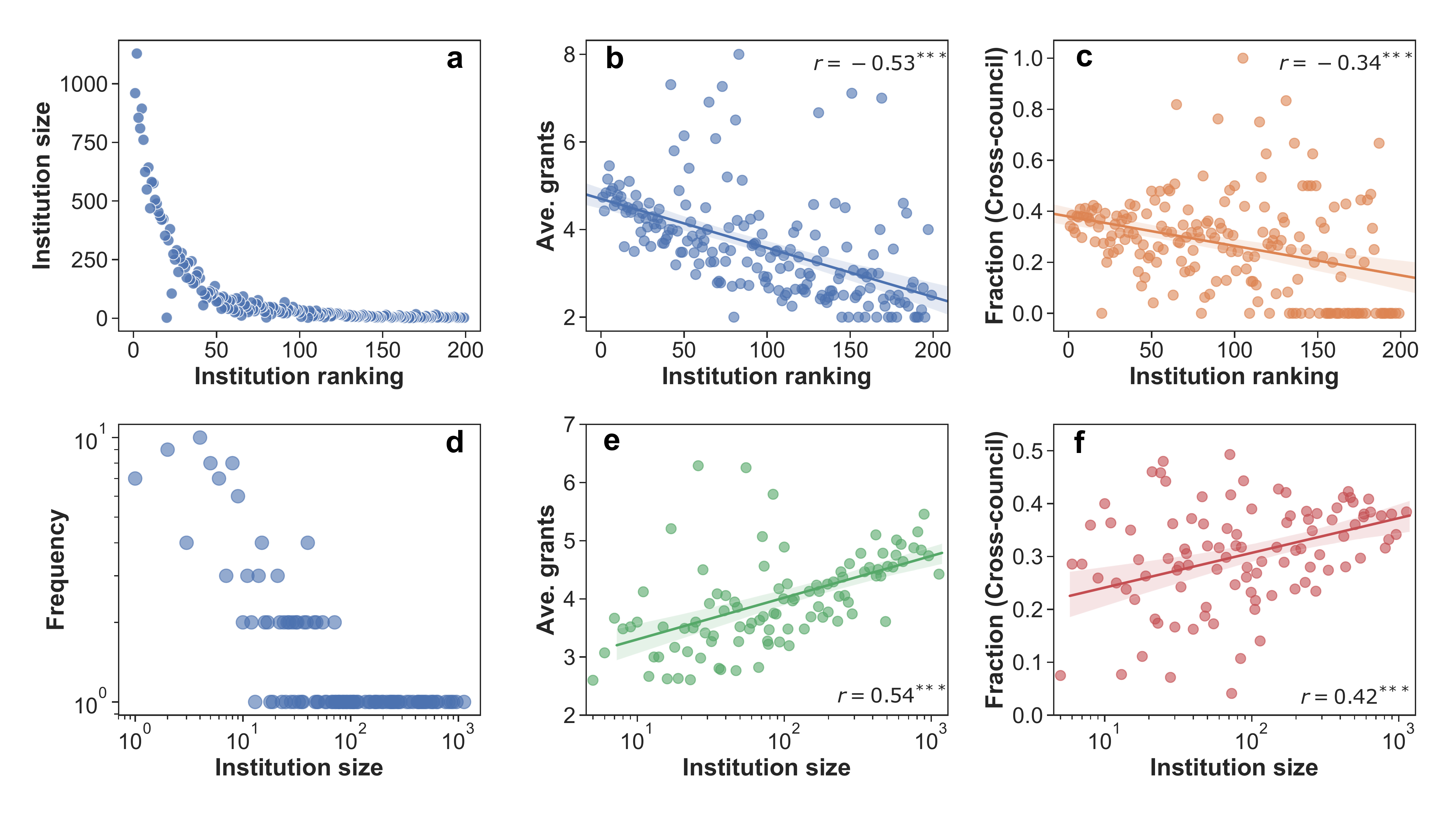}
    \caption{\textbf{Basic features of research institutions.} \textbf{(a)} The institution size against the institutional ranking, where the institution size is determined by the total number of distinct investigators that obtained at least one research grant under the institution, and the institution ranking is measured by the total amount of research grants awarded to an institution between $2006$ and $2018$. \textbf{(b)} The average number of grants obtained by each investigator generally decreased with the ranking of one's affiliation. \textbf{(c)} The correlation between institution ranking and the fraction of cross-council investigators is reported. \textbf{(d)} The institution size and its frequency. \textbf{(e)} The average number of grants obtained by each investigator positively correlates with the ranking of one's affiliation, indicating that the investigators in large institutions are likely to obtain more research grants on average. \textbf{(f)} A positive correlation between institution size and the fraction of cross-council investigators. The statistical model used to estimate a linear regression in \textbf{e-f} is in the form of $y \sim log(x)$. The solid line and the shaded area represent the regression line and the $95\%$ confidence interval, respectively. Each regression has also been annotated with the corresponding Pearson’s $r$ with $*** p<0.001$.}
    \label{fig:InstitutionFeature}
\end{figure*}

To examine the characteristics of institutions, we quantified the institution size by the total number of distinct investigators who have obtained at least one research grant under the institution, and institution ranking by the total amount of research grants awarded to an institution between $2006$ and $2018$. Institutions with less than $5$ funded research projects during the above period have been excluded, and there are a total of $199$ considered research institutions. We found that the number of investigators is very heterogeneously distributed among institutions (Fig.~\ref{fig:InstitutionFeature}a), 
with elite institutions having a higher number of funded investigators and most other institutions having very few funded investigators. Fig.~\ref{fig:InstitutionFeature}b reveals that the average number of research grants received per investigator is negatively correlated with one's institutional ranking. Fig.~\ref{fig:InstitutionFeature}c further illustrates that there is a negative correlation between the institutional ranking and the fraction of cross-council investigators, indicating that the investigators in elite institutions are more likely to secure funding from different research councils. This is probably because the leading institutions have stronger capacity to conduct research across various scientific domains.

Similarly, from the perspective of institution size, we found that most institution sizes are small, with only a few elite institutions with a high reputation having a large number of funded researchers (Fig.~\ref{fig:InstitutionFeature}d). Also, both the average number of grants obtained by each investigator and the fraction of cross-council investigators positively correlate with the size of their affiliations (Fig.~\ref{fig:InstitutionFeature}e-f), indicating that the investigators in large institutions averagely have obtained more research grants, and have more opportunities and resources to secure funding from different research councils.

\section{Robustness check for different window lengths and institution stratification} \label{sec:institution_trend}
\begin{figure*}
\centering
    \includegraphics[width=16cm]{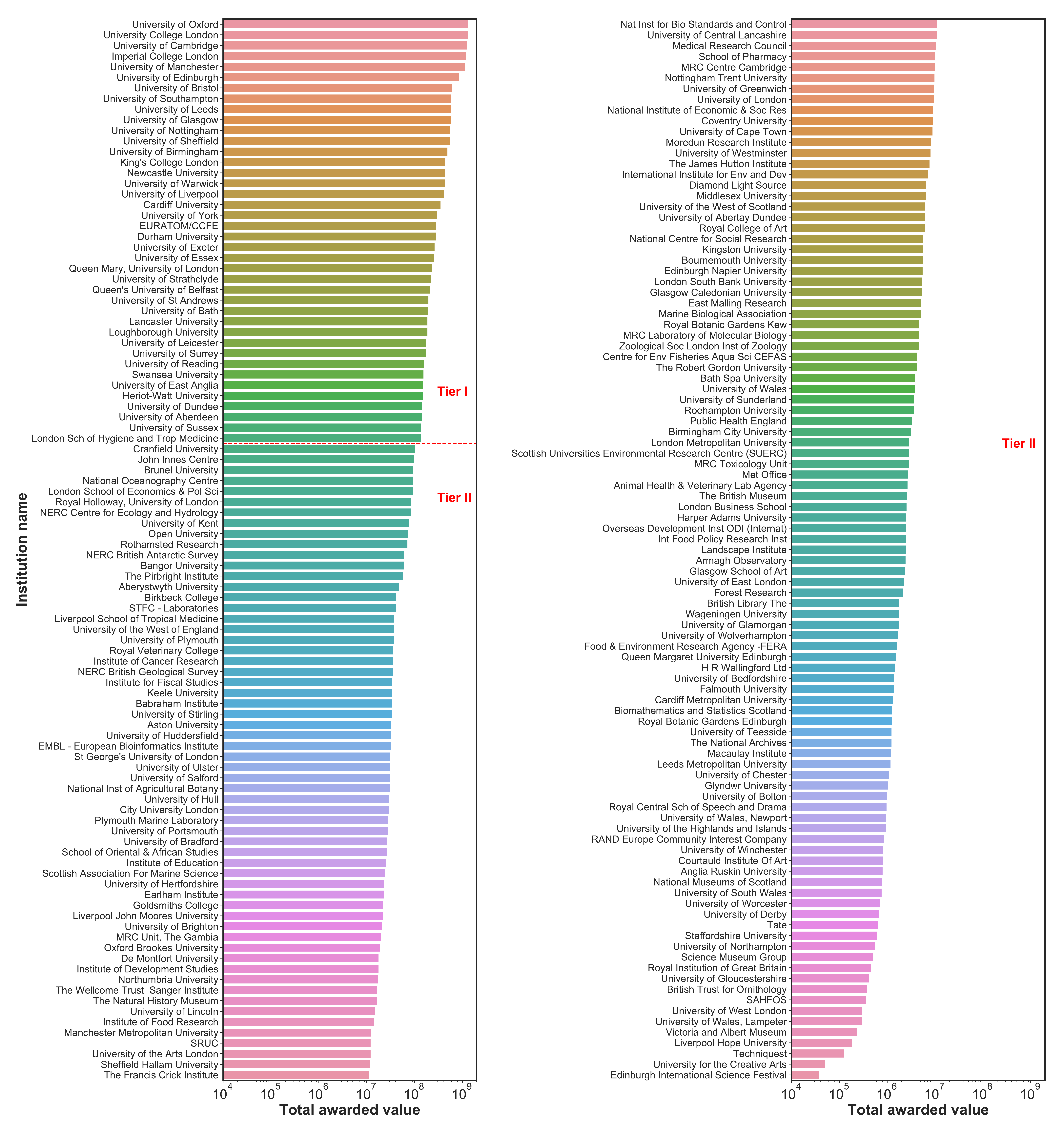}
    \caption{\textbf{Institutional ranking and stratification.} The research institutions are ranked according to their total awarded funding amount from $2006$ to $2018$. Here, we consider $199$ research institutions who have received at least $5$ research grants from the seven national research councils. The institutions are stratified into two tiers by checking whether their total awarded funding value is larger than the average amount per institution (i.e., $\textsterling 1.02 \times 10^8$ marked with dashed red line).}
    \label{fig:InstitutionRank}
\end{figure*}

\begin{figure*}
\centering
    \includegraphics[width=14cm]{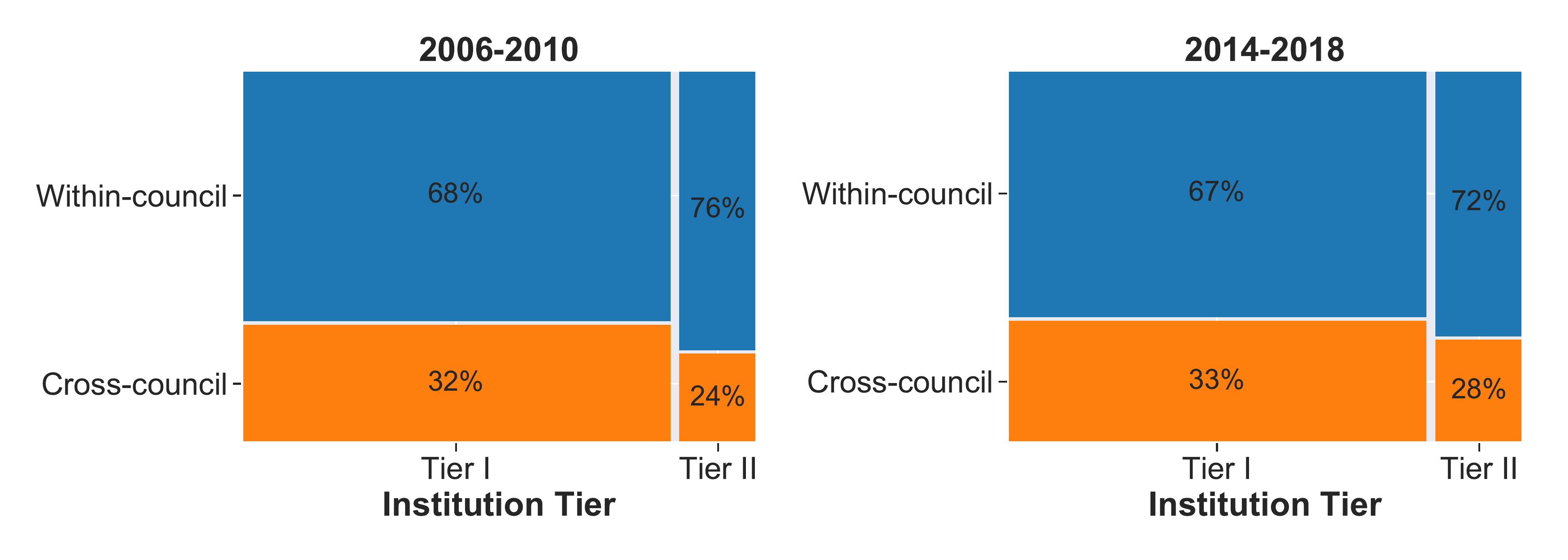}
    \caption{\textbf{The percentage of cross-council investigators in two compared 5-year time windows.} Here, the research institutions are stratified into two tiers by checking whether their total awarded funding is larger than the average amount per institution (i.e., $1.02 \times 10^8$). Box widths are proportional to the number of investigators in Tier I and Tier II, respectively. Box heights are proportional to the percentage of cross-council and within-council investigators. The institutions in Tier I have higher proportions of cross-council investigators than that in Tier II in both 5-y time windows ($\chi^2$ test $p<0.0001$, odds ratio$ = 1.49$ for 2006-2010; $p<0.0001$, odds ratio$ = 1.28$ for 2014-2018).}
    \label{fig:DifferenceTier2_5y}
\end{figure*}

To ensure that our results are not affected by selection of time window length, here we considered two 5-y compared period and repeated all our analysis. For the two tiers of institutions (Fig.~\ref{fig:InstitutionRank}), we found that the institutions in Tier I have higher proportion of cross-council investigators than that in Tier II in both 5-y time windows (Fig.~\ref{fig:DifferenceTier2_5y}), suggesting that the investigators in top institutions are generally more likely to secure research funding from different research councils. On the other hand, it is shown that Tier II institutions have a relatively higher growth rate in the fraction of cross-council investigators (with an increase from $24\%$ to $28\%$), which is four times greater than that in Tier I institutions (from $32\%$ to $33\%$). 

\begin{figure*}
\centering
    \includegraphics[width=14cm]{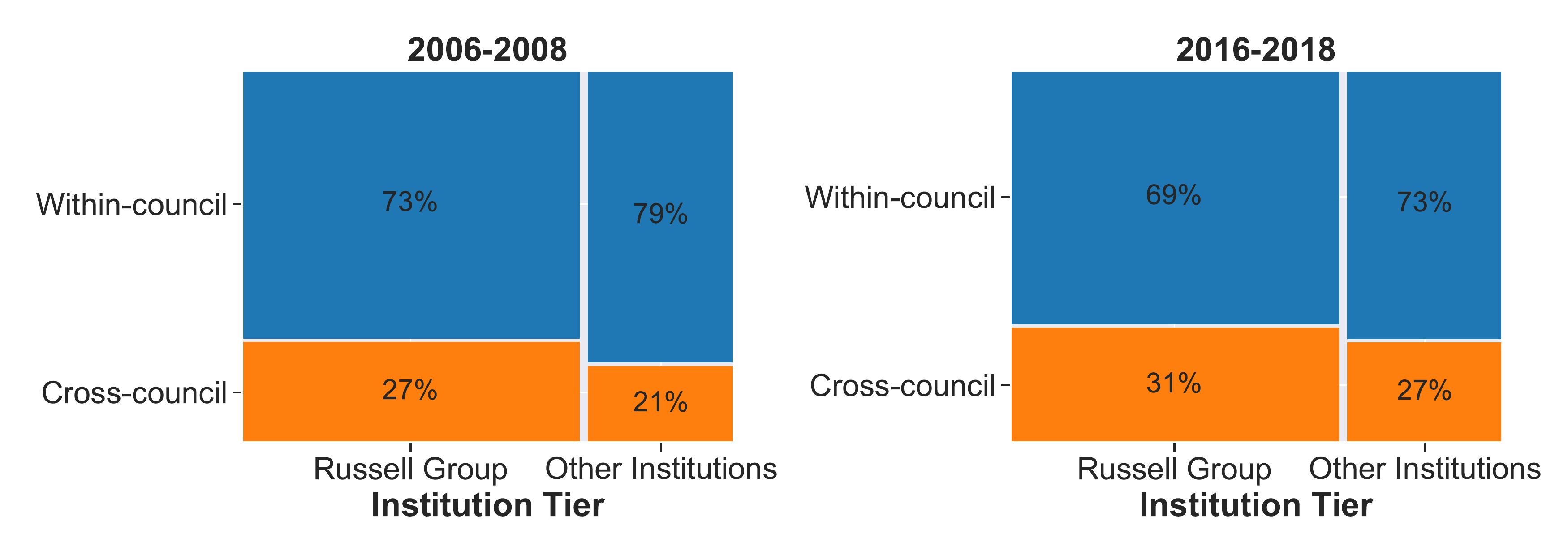}
    \caption{\textbf{The percentage of cross-council investigators in Russell group and other institutions.} Here, the research institutions are stratified into two groups: the Russell Group and other institutions. Box widths are proportional to the number of investigators in Tier I and Tier II, respectively. Box heights are proportional to the percentage of cross-council and within-council investigators. The institutions in Russell group have higher percentages of cross-council investigators than that in the other institution group in both 3-y time windows ($\chi^2$ test $p<0.0001$, odds ratio$ = 1.44$ for 2006-2010; $p<0.0001$, odds ratio$ = 1.21$ for 2016-2018).}
    \label{fig:DifferenceRussell_3y}
\end{figure*}

Moreover, to test if the results would be affected by the approach of institution stratification, we also tried to classify the institutions into two groups based on whether they belong to the Russell Group. In total, the Russell Group includes $24$ top universities in the UK, all of which have a strong reputation for academic achievement. We can see that in both two 3-y periods, the proportions of cross-council investigators in Russell Group are relatively higher than that in the group of other institutions  (Fig.~\ref{fig:DifferenceRussell_3y}). Similarly, we also found that the group of other institutions exhibits a higher growth rate, by a factor of $28.6\%$, while the Russell Group represents a lower growth rate, with an increase of $14.8\%$. 

\section{Evolution of collaboration networks and consistent structural advantage for cross-council investigators}\label{sec:structral_advantage}
\begin{figure*}
\centering
    \includegraphics[width=16cm]{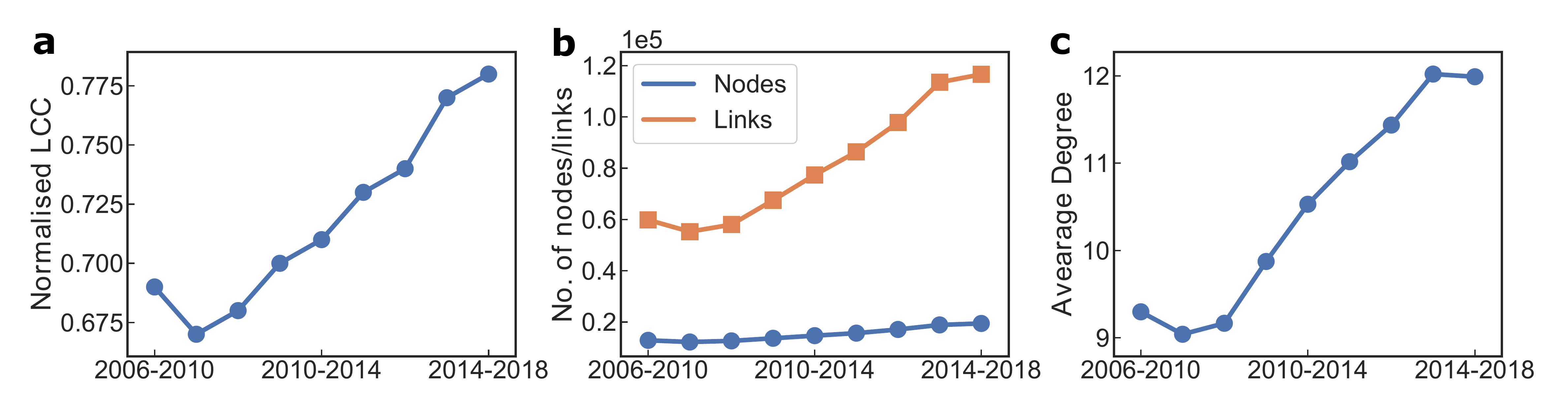}
    \caption{\textbf{Network properties of the collaboration network for each 5-y sliding time window between $2006$ and $2018$.} (a) The normalized LCC grows over time, indicating the collaboration network becomes more interconnected. The normalized LCC is calculated by dividing the size of the LCC by the total number of nodes in the network. (b) The number of nodes and links in the collaboration network. (c) The average degrees of collaboration network increase over time.}
    \label{fig:LCC}
\end{figure*}

To examine the evolution of the collaboration network, we used a sliding window of five years within which we aggregated all the partnerships occurring in the research projects. Starting in $2006$, we shifted the sliding the window in one-year increments, and obtained a total of $9$ time slices between $2006$ and $2018$. We found that the normalized size of the largest connected component (LCC) increased steadily with time (Fig.~\ref{fig:LCC}a), reflecting better interconnectivity among different local scientific communities in the funding landscape. Note that the normalized size of LCC is obtained by dividing the size of the LCC by the total number of nodes. Moreover, the number of links in the collaboration network of each time slice grows much faster than the number of nodes (Fig.~\ref{fig:LCC}b). The average degree of collaboration networks shows a significant increase over periods, indicating that the investigators tend to build a much broader collaboration(Fig.~\ref{fig:LCC}c).

Furthermore, we checked whether the structural advantage of cross-council investigators exists in different time windows. We constructed a collaboration network and examined the network using two 5-y time windows, namely the first 5-y window (2006-2010) and the last 5-y window (2014-2018), respectively. There are $69\%$ of the investigators connected in the LCC in 2006-2010, and $78\%$ of that in 2014-2018. Through the network analysis in the LCCs, we calculated four node-level metrics, namely, degree centrality, closeness centrality, betweenness centrality and normalized effective size (see Materials and Methods). The definitions of three representative centrality measures are characterised as follows:
\begin{enumerate}[label=(\roman*)]
\item \emph{Degree centrality}. The normalized degree~\cite{wasserman1994social} centrality of node $i$ is defined as:
\begin{equation}
\label{eq:degree}
dc_{i}= \frac{k_{i}}{N-1}
\end{equation}
where $k_{i}$ is the degree (number of links) of node $i$. The degree centrality values are normalized by dividing by the maximum possible degree $N-1$ in the LCC where $N$ is the total number of nodes in LCC.

\item \emph{Closeness centrality.} The closeness centrality~\cite{wasserman1994social} of a node $i$ is defined as the reciprocal of the average shortest path distance to $i$ over all $N-1$ reachable nodes in the LCC:
\begin{equation}
\label{eq:clossness}
c_{i}=\frac{N-1}{\sum^{N-1}_{j=1}d(i, j)}
\end{equation}
where $d(i, j)$ is the shortest-path distance between $i$ and $i$, and $N-1$ is the number of nodes that can reach $i$. Closeness centrality of a node reflects the overall distance between that node and the rest of the nodes in the network.

\item \emph{Betweenness centrality.} The betweenness centrality~\cite{wasserman1994social} of node $i$ is defined as follows:
\begin{equation}
\label{eq:betweeness}
b_{i}=\frac{2}{(N-1)(N-2)}\sum\limits_{a,b \in V}\frac{n_{a,b}(i)}{n_{a,b}}
\end{equation}
where $V$ is the set of nodes in LCC, $n_{a,b}$ is the total number of shortest paths between nodes $a$ and $b$, and $n_{a,b}(i)$ is the number of shortest paths between $a$ and $b$ that actually go through $i$. If $a=b$,$n_{a,b}=1$, and if $i \in \{a,b\}, n_{a,b}(i)=0$.
\end{enumerate}

\begin{figure*}
\centering
    \includegraphics[width=16cm]{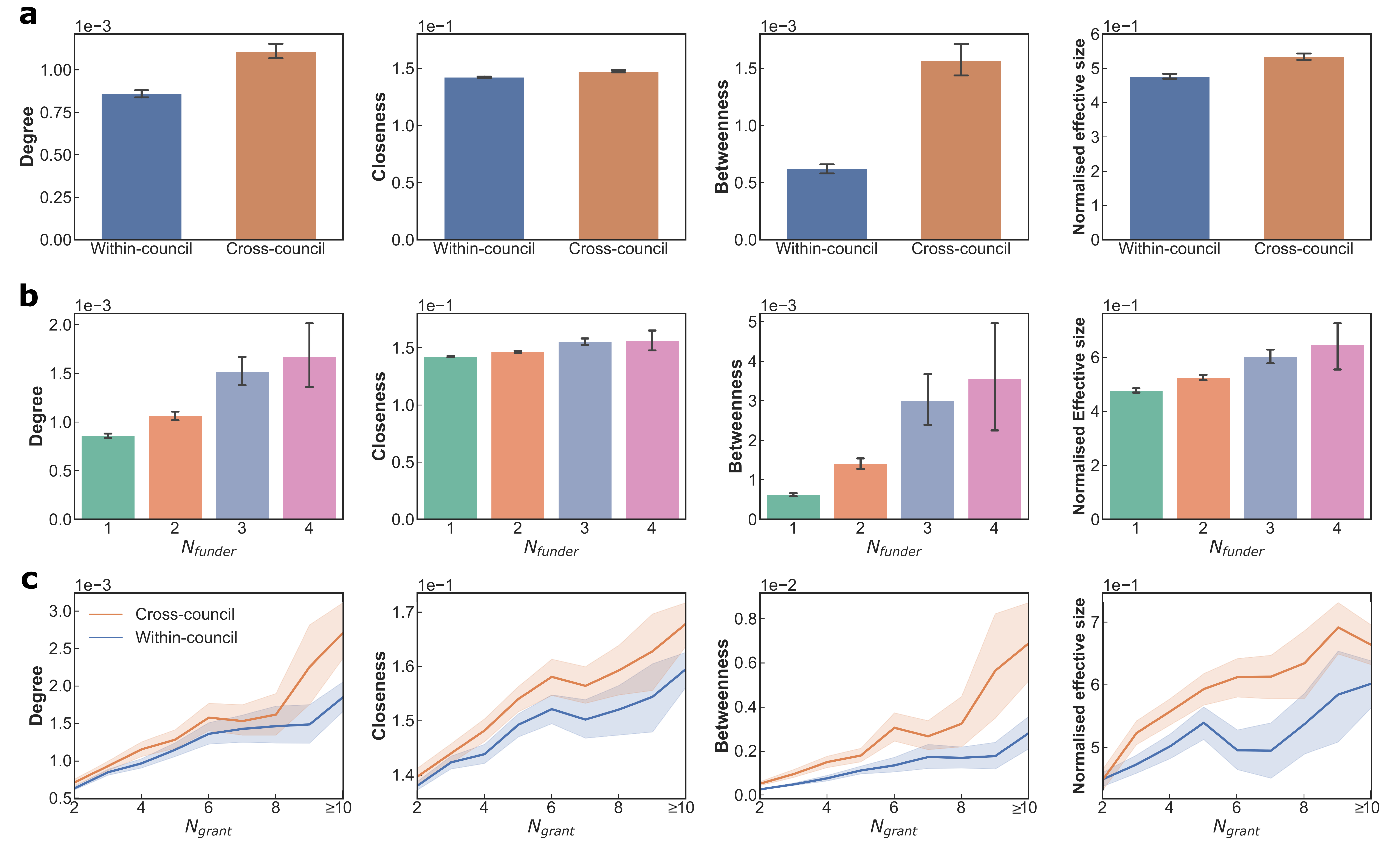} 
    \caption{\textbf{Network structural advantages of cross-council investigators between $2006$ and $2010$.} Each column corresponds to a different network property, namely (from left to right): the degree centrality, closeness centrality, betweenness and normalized effective size. \textbf{(a)} Cross-council investigators significantly outperform the within-council investigators in four network properties (Welch’s t-test $p<0.001$). \textbf{(b)} Network metrics among investigators increase with the number of councils they have received funding from. \textbf{(c)} Cross-council investigators consistently outperform within-council investigators, regardless of the number of grants they have obtained in the period. The same applies in the case of the degree only for investigators who receive large numbers of grants. The error bars and shaded areas both represent $95\%$ confidence intervals.}
    \label{fig:NetworkCompare2006}
\end{figure*}

\begin{figure*}
\centering
    \includegraphics[width=16cm]{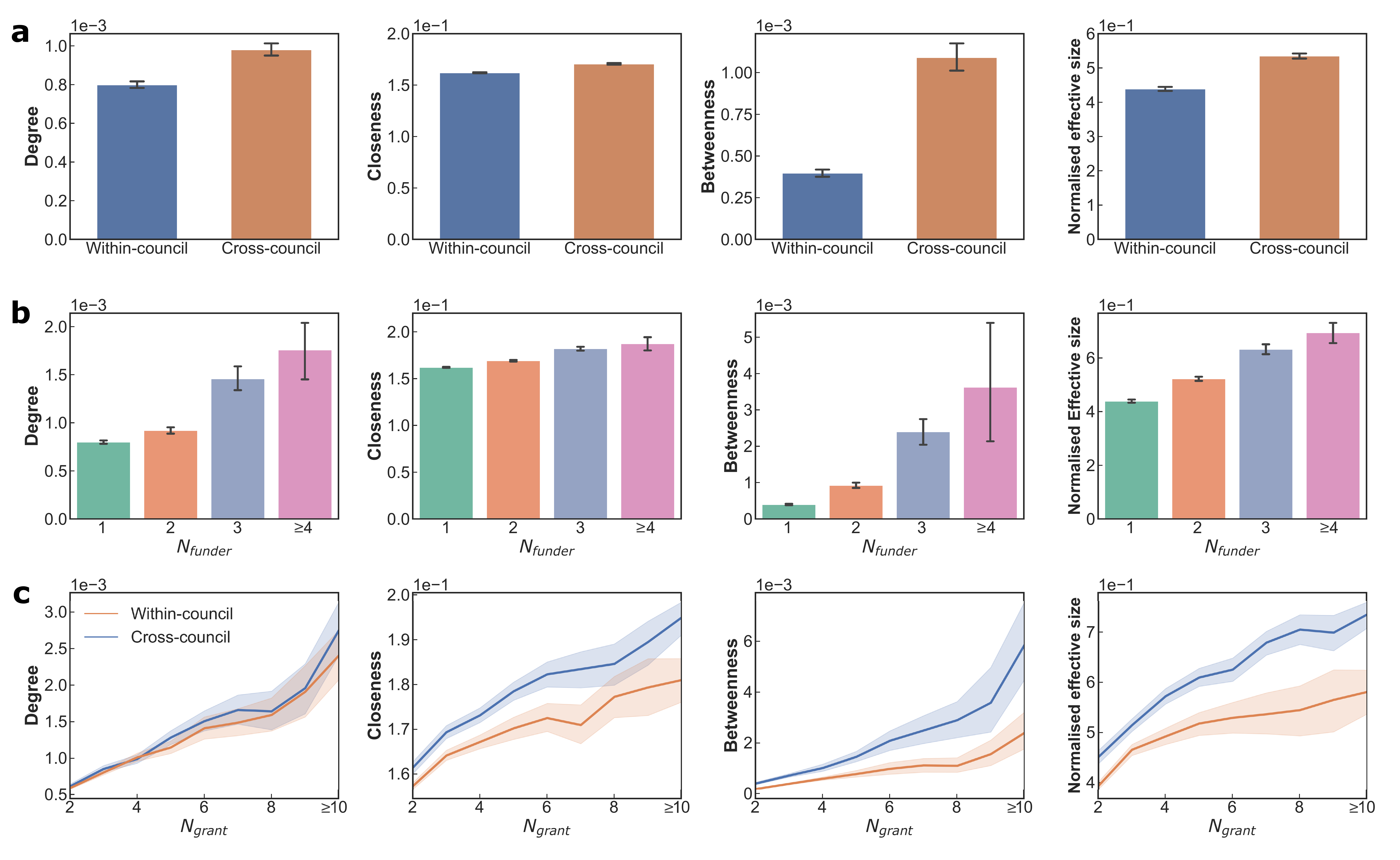}
    \caption{\textbf{Network structural advantages of cross-council investigators between $2014$ and $2018$.} Each column corresponds to a different type of network properties, namely (from left to right): the degree centrality, closeness centrality, betweenness and normalized effective size. \textbf{(a)} Cross-council investigators significantly outperform the within-council investigators in four network properties (Welch’s t-test $p<0.001$). \textbf{(b)} Network metrics among investigators increase with the number of councils they have received funding from. \textbf{(c)} Cross-council investigators consistently outperform within-council investigators in terms of closeness, betweenness and normalized effective size, regardless of the number of grants they have obtained in the period. The error bars and shaded areas both represent $95\%$ confidence intervals.}
    \label{fig:NetworkCompare2014}
\end{figure*}

We then quantitatively compared the structural position difference between the cross-council and within-council investigators in terms of these four metrics. In general, we found that, in both two compared periods, cross-council investigators consistently outperform the within-council investigators in terms of network centrality and brokerage power, where the conclusions remain the same as in the main text (Fig.\ref{fig:NetworkCompare2006} and Fig.\ref{fig:NetworkCompare2014}).

\section{Quantifying PIs' funding performance and outcomes}\label{sec:grant_pi_feature}
\subsection{Grant-related features and their adjustments across councils and years}\label{sec:grant_feature}
In this section, we enumerated the six attributes of research grants that are considered in our study, followed by the percentile adjustments of these attributes over discipline-based research councils and years. Here, we introduce the following notations:

\begin{enumerate}
\item \textbf{$V(g_{i})$}: The funded value of grant $g_{i}$.

\item \textbf{$D(g_{i})$}: The time duration of grant $g_{i}$.

\item \textbf{$T(g_{i})$}: The number of investigators in grant $g_{i}$.

\item \textbf{$N_{p}(g_i)$}: The number of papers funded by the grant $g_{i}$:
\begin{equation}
N_{p}(g_i) = |Papers(g_{i})|
\end{equation}
where $Papers(g_{i})$ denotes the set of papers that funded by grant $g_{i}$.

\item \textbf{$C_5(g_{i})$}: The total normalized citations that grant $g_{i}$ accumulates from the funded papers $5$ years after its publication:
\begin{equation}
\centering
C_5(g_{i}) = \sum_{p_{j} \in Papers(g_{i})} C_5(p_{j})
\end{equation}
where $C_5(p_{j})$ represents the normalized citations that paper $p_{j}$ accumulates $5$ years after publication. 

\item \textbf{$\langle C_5(g_{i}) \rangle$}: The average normalized citations that grant $g_{i}$ accumulates from the funded papers $5$ years after its publication:
\begin{equation}
\centering
\langle C_5(g_{i}) \rangle = \frac{\sum\limits_{p_{j} \in Papers(g_{i})} C_5(p_{j})}{N_{p}(g_i)}
\end{equation}
\end{enumerate}

Many studies have shown that: (i) the number of received citations fluctuate greatly in different disciplines, and (ii) the average number of citations per paper changes over time~\cite{radicchi2008universality,althouse2009differences,sinatra2016quantifying}. Thus, we here followed the authoritative approach~\cite{radicchi2008universality} and normalized the citations of each paper by the average citations of all papers belonging to the same year and discipline (including $19$ scientific disciplines in Microsoft Academic Graph (MAG) dataset).

\begin{figure*}
\centering
    \includegraphics[width=16cm]{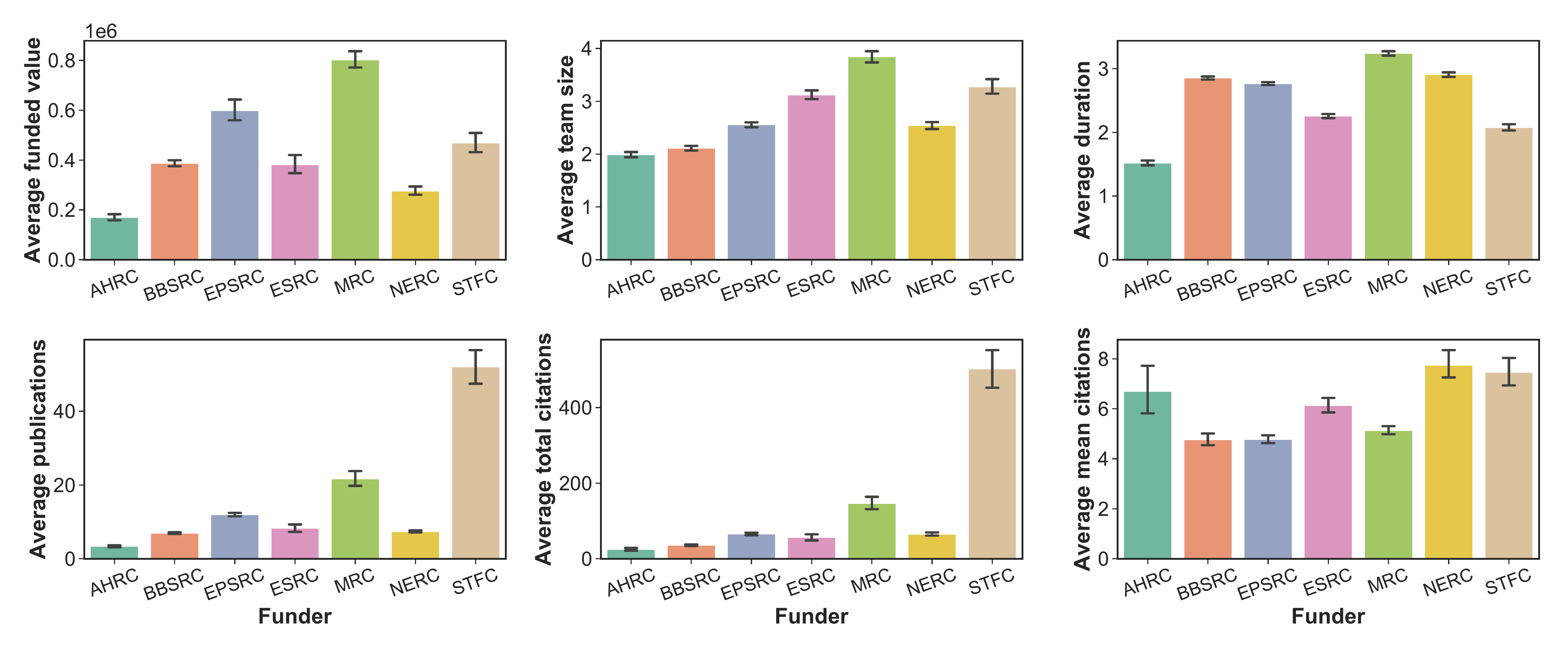}
    \caption{\textbf{Comparison of project properties and outcomes between the seven research councils.} We compared six different characteristics of research projects as follows: (1) the average funded value per grant; (2) the average team size across the awarded grants; (3) the average duration per project; (4) the average number of publications reported per project; (5) the average total citations of papers received within 5 years of publication across funded projects; and (6) the average mean citations of papers received within 5 years of publication across funded projects. The attributes of research projects vary greatly among different councils, especially in terms of average funded value and average number of reported publications. Error bar represents the 95\% confidence interval.}
    \label{fig:DifferenceCouncil}
\end{figure*}

Fig.~\ref{fig:DifferenceCouncil} illustrates that the six attributes of research projects vary greatly among different councils, especially in terms of average funded value and average number of reported publications. For instance, the average number of publications per grant in $STFC$ is nearly fifteen times larger than that in $AHRC$. In order to fairly compare grant properties and research outcomes across different research council and year, we adjusted these features by calculating their percentile rank relative to the values in the same council and year. More formally, let $P(X(g_i))$ denote the percentile rank of a considered feature value of grant $g_i$ relative to a list of grants in the same council and year, where $X \in \{ V,D,T, N_{p}, C_5, \langle C_5 \rangle \}$. For example, $P(V(g_i)) = 80$ means that $80\%$ of the observed grant value are below the given value of grant $g_i$; Similarly, $P(T(g_i)) = 30$ indicates that $30\%$ of grant team size are smaller than that of the given grant $g_i$. In this way, the attributes of research grants awarded by different research councils and years are comparable.

\subsection{PIs' Funding performance and research outcomes} \label{sec:PI_feature}
After obtaining the features of research grants, we then quantify the funding performance and research outcomes for any given principal investigators $s_k$ based on their involved grants in an observing period. Here, we introduce the following notations:
\begin{enumerate}
\item \textbf{$N_{g}(s_k)$}: The number of grants obtained by principal investigator $s_{k}$: 
\begin{equation}
N_{g}(s_k) = |Grants(s_k)|
\end{equation}
where $Grants(s_{k})$ denotes the set of research grants that obtained by investigator $s_{k}$ in a given time period.

\item \textbf{$R(s_k)$}: The institution ranking of principal investigator $s_{k}$. Here, we measure institutions' ranking based on their total awarded amounts of funding from $2006$ and $2018$. We only consider the institutions at least receiving $5$ research grants during the period, and there are a total of $199$ scientific institutions.

\item \textbf{$\langle P(X(g_{i})) \rangle_{s_{k}}$}: The average percentile features of grants obtained by principal investigator $s_k$:
\begin{equation}
\langle P(X(g_{i})) \rangle_{s_{k}} = \frac{\sum\limits_{g_i \in Grants(s_k)} P(X(g_i))}{N_{g}(s_k)}
\end{equation}
where $X \in \{ V,D,T, N_{p}, C_5, \langle C_5 \rangle \}$, referring to the six attributes of research grants mentioned above. For example, $\langle P(V(g_{i})) \rangle_{s_{k}}$ indicates the average percentile funded value across research grants received by principal investigator $s_k$. Similarly, the average percentile of total normalized citations per project led by investigator $s_k$ is defined as $\langle P(C_5(g_{i})) \rangle_{s_{k}}$.
\end{enumerate}
Based on these metrics, we compared research outcomes and performance in funding between each paired investigators.

\newpage

\begin{table*}[bt!]
\centering
\begin{threeparttable}
\caption{\textbf{Summary of research projects in seven research councils between $2006$ and $2018$.} $N_{g}$ represents the number of research grants. $V$ indicates the total amount of funding allocated to each council. $N_{inv}$, $N_{ins}$ and $N_{sub}$ denote the number of different investigators, research institutions and subjects, respectively.} \label{tab:CouncilProperty}
{\small
\begin{tabular}{p{8.8cm}p{1cm}p{1cm}<{\centering}p{1.2cm}<{\centering}p{0.8cm}<{\centering}p{1cm}<{\centering}p{0.6cm}<{\centering}}
\hline
\hline
    Research Council & Abbr. & $N_{g}$ & $V (\times 10^{9})$ & $N_{inv}$ & $N_{ins}$ & $N_{sub}$\\
    \hline
    Arts and Humanities Research Council & AHRC & $4,488$ & $0.8$ & $6,160$  & $158$ & $63$\\
    Biotechnology and Biological Sciences Research Council & BBSRC & $7,878$ & $3.0$  &  $6,573$ & $149$ &  $54$\\
    Engineering and Physical Sciences Research Council & EPSRC & $13,712$ & $8.2$  & $13,332$ & $153$ & $68$\\
    Economic and Social Research Council & ESRC & $4,910$ & $1.9$ & $9,849$  & $187$ & $62$\\
    Medical Research Council & MRC & $4,507$  & $3.6$  &   $10,004$  & $162$ & $22$\\
    Natural Environment Research Council & NERC & $5,879$ & $1.6$ & $5,398$  &  $197$ & $71$\\
    Science and Technology Facilities Council& STFC & $3,045$ &  $1.4$  &  $2,428$  &$172$ & $58$\\
\bottomrule
\end{tabular}}
\end{threeparttable}
\end{table*}

\clearpage

\clearpage
\begin{table*}[h!]
\centering
\begin{threeparttable}
\caption{\textbf{Regression analysis of cross-council behavior on research outcomes and scientific impact.} Along with each indicator of outputs, the following predictor variables were used: PI's institutional ranking, the number of awarded grants and their average value, team size and duration. The regression models are computed in matched datasets. Standard errors for the coefficients are reported in parenthesis. *$p<0.1$; **$p<0.05$; ***$p<0.01$} 
\begin{tabular}{p{4cm}p{3cm}<{\centering}p{3cm}<{\centering}p{3cm}<{\centering}}
\hline
\hline
&\multicolumn{3}{c}{Dependent variables: research outcomes and scientific impact}\\
\cmidrule(lr){2-4} 
& Average papers  & Average total citations  & Average mean citations\\
\hline
\bf{Cross-council}  & $-1.92^{**}$ & $-3.37^{***}$ & $-3.65^{***}$ \\
                    & $(0.86)$     & $(0.86)$      & $(1.00)$\\
Institution ranking & $-0.03^{*}$  & $-0.04^{**}$  & $-0.04^{*}$\\
                    & $(0.02)$     & $(0.02)$      & $(0.02)$\\
Number of grants    & $0.25$       & $0.42$        & $0.39$  \\
                    & $(0.24)$     & $(0.26)$      & $(0.28)$ \\
Average funded value& $0.30^{***}$ & $0.29^{***}$  & $0.15^{***}$\\
                    & $(0.03)$     & $(0.03)$      & $(0.03)$ \\
Average team size   & $0.05^{**}$  & $0.07^{***}$  & $0.04$\\
                    & $(0.02)$     & $(0.02)$      & $(0.03)$\\
Average duration    & $0.19^{***}$ & $0.12^{***}$  & $-0.01$\\
                    & $(0.03)$     & $(0.03)$      & $(0.03)$\\
Constant            & $24.50^{***}$& $28.77^{***}$ & $42.95^{***}$\\
                    & $(1.76)$     & $(1.87)$      & $(2.03)$\\
\hline
\hline
No. of pairs        & $958$       & $958$          & $958$  \\
Adjusted $R^2$      & $\bf{0.21}$  & $\bf{0.15}$   & $\bf{0.03}$\\
\bottomrule
\end{tabular}
\label{tab:RegressionOutcome}
\end{threeparttable}
\end{table*}

\begin{table*}[h!]
\centering
\begin{threeparttable}
\caption{\textbf{Regression analysis of cross-council behavior on long-term funding performance.} Along with each dependent variable of funding performance, the following dependent variables were used: PI's institutional ranking, the number of awarded grants and their average value, team size, duration, the number of publications, total citations and mean citations. The regression models are computed in matched datasets. Standard errors for the coefficients are reported in parenthesis. *$p<0.1$; **$p<0.05$; ***$p<0.01$}
\begin{tabular}{p{4cm}p{2.5cm}<{\centering}p{2.5cm}<{\centering}p{2.5cm}<{\centering}p{2cm}<{\centering}}
\hline
\hline
&\multicolumn{4}{c}{Dependent variables: funding performance in out-of-sample period}\\
\cmidrule(lr){2-5}
& Number of grants  & Average funded value  & Average team size & Average duration \\
\hline
\bf{Cross-council}  & $0.23^{**}$ & $3.35^{**}$   & $3.68^{**}$   & $1.24$ \\
                    & $(0.10)$    & $(1.60)$      & $(1.56)$      & $(1.62)$      \\
Institution ranking & $-0.003$     & $-0.07^{**}$   & $-0.01$     & $-0.04$  \\
                    & $(0.002)$   & $(0.03)$      & $(0.03)$      & $(0.03)$       \\
Number of grants    & $0.60^{***}$& $4.19^{***}$  & $4.04^{***}$  & $4.15^{***}$\\
                    & $(0.04)$    & $(0.61)$      & $(0.59)$      & $(0.61)$        \\
Average funded value& $0.001$     & $0.21^{***}$  & $-0.03$       & $0.02$ \\
                    & $(0.004)$   & $(0.06)$      & $(0.05)$      & $(0.06)$       \\
Average team size   & $0.002$     & $0.11^{**}$   & $0.32^{***}$  & $0.06$  \\
                    & $(0.002)$   & $(0.04)$      & $(0.04)$      & $(0.04)$       \\
Average duration    & $0.01^{*}$  & $-0.0007$     & $0.05$        & $0.11^{**}$  \\
                    & $(0.003)$   & $(0.05)$      & $(0.05)$      & $(0.05)$       \\
Average publications & $-0.01^{*}$& $0.10$        & $0.05$        & $0.13$ \\
                    & $(0.01)$    & $(0.13)$      & $(0.13)$      & $(0.13)$       \\
Average total citations & $0.03^{**}$& $0.17$     & $0.24$        & $0.16$ \\
                    & $(0.01)$    & $(0.21)$      & $(0.21)$      & $(0.21)$        \\
Average mean citations & $-0.02^{*}$ & $-0.11$    & $-0.15$       & $-0.08$  \\
                    & $(0.01)$    & $(0.12)$      & $(0.12)$      & $(0.12)$       \\
Constant            & $-0.70^{***}$& $-0.07$      & $-3.06$        & $3.90$  \\
                    & $(0.25)$    & $(3.85)$      & $(3.75)$      & $(3.88)$       \\
\hline
\hline
No. of pairs        & $709$       & $709$      & $709$     & $709$ \\
Adjusted $R^2$      & $\bf{0.17}$   & $\bf{0.10}$   & $\bf{0.10}$   & $\bf{0.08}$  \\
\bottomrule
\end{tabular}
\label{tab:RegressionFuture}
\end{threeparttable}
\end{table*}

\clearpage

\end{document}